\documentclass[%
 reprint,
preprintnumbers,
 amsmath,amssymb,
 aps,
 prd,
]{revtex4-2}

\usepackage{graphicx}
\usepackage{dcolumn}
\usepackage{bm}
\usepackage[
    colorlinks=true,
    linkcolor=blue,
    citecolor=blue,
    urlcolor=blue
]{hyperref}
\usepackage{multirow}
\usepackage{color}

\begin{document}

\preprint{UTHEP-813, UTCCS-P-171}

\title{
Sparse modeling study of extracting charmonium spectral functions
from lattice QCD at finite temperature
}

\author{Junichi Takahashi}
 \affiliation{
 Meteorological College, Japan Meteorological Agency,
 7-4-81, Asahi-cho, Kashiwa, Chiba 277-0852, Japan
 }
 \email{
 mhjkk-takahashi@met.kishou.go.jp
 }
 
\author{Hiroshi Ohno}%
 \affiliation{%
 Center for Computational Sciences, University of Tsukuba,
 1-1-1, Tennodai, Tsukuba, Ibaraki 305-8577, Japan
 }
 \email{
 hohno@ccs.tsukuba.ac.jp
 }

\author{Akio Tomiya}
 \affiliation{
 Department of Information and Mathematical Sciences, Division of Mathematical Sciences, Tokyo Women's Christian University,
 2-6-1 Zempukuji, Suginami-ku, Tokyo 167-8585, Japan
 }
 \affiliation{RIKEN Center for Computational Science, Kobe 650-0047, Japan} 
 \affiliation{
 Department of Physics, Kyoto University, Kyoto 606-8502, Japan}
 \email{
 akio@yukawa.kyoto-u.ac.jp
 }

\date{\today}

\begin{abstract}
We present charmonium spectral functions extracted from Euclidean-time correlation functions using sparse modeling (SpM). SpM solves inverse problems by considering only the sparsity of the target solution. To assess the applicability of the method, we first test it with mock data designed to mimic charmonium correlation functions. We demonstrate that while resonance peaks in the spectral functions can be reconstructed using this method, transport peaks are difficult to resolve without introducing further assumptions beyond sparsity. We then apply the method to charmonium correlation functions obtained from lattice QCD at temperatures below and above the critical temperature. The results are found to be qualitatively consistent with those obtained using the maximum entropy method, although the transport peak is not clearly resolved.
This indicates that, even when relying solely on the assumption of sparsity, the method can capture some relevant features of the underlying physics.

\end{abstract}

\maketitle


\section{
Introduction
\label{sec:Introduction}
}
Meson spectral functions provide essential information on the properties of hot and dense
QCD matter created in relativistic heavy-ion collisions. They encode the real-time dynamics
of strongly interacting systems and are directly related to phenomenologically relevant
observables. For example, the vector spectral function determines the thermal dilepton
production rate~\cite{McLerran:1984ay, Braaten1990.PhysRevLett.64.2242, moore2006dileptons},
which offers an experimentally accessible probe of the electromagnetic response of
the quark–gluon plasma (QGP).
\\
\indent
Quarkonium, a bound state of a heavy quark and antiquark, serves as a valuable probe of
QGP created in relativistic heavy-ion collisions~\cite{Matsui:1986dk}.
Due to the large quark mass, quarkonium systems can be described within a nonrelativistic framework,
which allows to relate their in-medium properties to an effective heavy-quark potential.
At finite temperature, this potential becomes complex-valued, reflecting the interaction of
the heavy quark–antiquark pair with the surrounding thermal medium~\cite{Laine:2006ns, Burnier:2007qm, Burnier:2014ssa, Bala:2019cqu, Ali:2025iux}.
The real part of the potential is modified by the presence of color charges in the deconfined medium,
leading to color Debye screening of the quark–antiquark interaction. As the temperature increases,
this screening weakens the binding of quarkonium states.
In addition, the imaginary part of the finite-temperature potential encodes medium-induced
dissociation processes, such as Landau damping and singlet–octet transitions caused by interactions
with thermal gluons. These effects give rise to a finite thermal width of quarkonium states, implying that
quarkonium melting should be understood not only as the disappearance of bound states due to screening,
but also as the loss of well-defined resonances caused by strong in-medium damping.
\\
\indent
Experimental studies at RHIC and LHC have provided compelling evidence for such
in-medium modifications of quarkonia.
Measurements of the $J/\Psi$ and $\Upsilon$ yields reveal significant suppression compared
to the proton-proton baselines~\cite{PHENIX:2006gsi, PHENIX:2014tbe, ALICE:2016flj,
ALICE:2018wzm, CMS:2018zza}. These measurements provide essential information on
the onset of deconfinement and the properties of the QGP, and motivate further
theoretical studies to understand the in-medium modification of quarkonium, requiring
knowledge of the corresponding spectral function.
Beyond quarkonium spectroscopy, the low-frequency structure of the vector spectral function
is directly related to transport coefficients such as the heavy quark diffusion
constant~\cite{Petreczky2006.PhysRevD.73.014508}, which are essential to describe transport
phenomena in heavy-ion experiments.
\\
\indent
In lattice QCD, however, spectral functions cannot be computed directly. Instead, they are
inferred from meson correlation functions in Euclidean time. Reconstructing spectral functions
from these correlation functions is, however, an ill-posed inverse problem: the limited number
of temporal data points and the presence of statistical noise make it difficult to uniquely
determine the spectral shape.

Various approaches have been proposed to address this difficulty,
including the maximum entropy method (MEM)~\cite{ASAKAWA2001459},
the Backus-Gilbert method~\cite{Brandt.PhysRevD.92.094510},
the Chybyshev polynomial method~\cite{Hashimoto:2025ohw},
and stochastic methods~\cite{Ding2018.PhysRevD.97.094503}.
Each of these has its strengths and limitations, and the results may depend sensitively
on algorithmic choices. For this reason, it is necessary to compare different reconstruction
methods in order to assess systematic uncertainties.

Recently, sparse modeling (SpM) has been introduced to the lattice QCD community as a promising
alternative for spectral reconstruction.
Originally developed in condensed matter physics~\cite{Otsuki.PhysRevE.95.061302, Shinaoka.PhysRevB.96.035147},
it enables stable reconstructions with reduced bias and controlled resolution,
while relying on fewer explicit model assumptions compared to conventional approaches.
In particular, Itou and Nagai~\cite{itou2020sparse} successfully applied SpM
to the correlation functions of the energy–momentum tensor, demonstrating its potential
to extract transport coefficients such as shear viscosity.
Motivated by the question of whether essential quarkonium physics can be extracted within a framework with minimal model assumptions,
we adopt SpM in this study.
Extending this approach to
the study of quarkonium correlators at finite temperature provides a new opportunity
to obtain more reliable information on in-medium spectral functions,
including quarkonium survival patterns and transport properties relevant for heavy-ion phenomenology.

In this work, we investigate the charmonium spectral function at finite temperature
using SpM. By performing analyses with both mock data and data from
lattice QCD calculations and by comparing our results with conventional methods,
we assess systematic uncertainties and establish the applicability of SpM
for quantitative studies of heavy-quark observables in the QGP.

The remainder of the paper is organized as follows.
In Sec.~\ref{sec:sparse_modeling},
we give a brief review of the SpM method,
explain our ideas and describe the procedure for using the method
to extract spectral functions in our study.
In Sec.~\ref{sec:mock_data_tests},
we explain the mock-data tests of the method
and discuss the results of that test.
Section~\ref{sec:results_LQCD} presents the results of the spectral functions
using actual lattice QCD data.
Finally we summarize our study in Sec.~\ref{sec:summary}.

\section{
Sparse Modeling
\label{sec:sparse_modeling}
}
In this section, we introduce the SpM method. In Sec.~\ref{subsec:review_SpM},
we briefly review its formulation in the intermediate representation (IR) basis,
following Refs.~\cite{Otsuki.PhysRevE.95.061302,Ohtsuki_JPSJ.89.012001}.
While previous studies have generally neglected the covariance between
correlation functions at different Euclidean times, SpM is capable of incorporating
it~\cite{Ohtsuki_JPSJ.89.012001}, and we explicitly take this covariance into account
in our analysis, as explained in Sec.~\ref{subsec:consider_C_for_SpM}.
Finally, because the method is employed here to extract spectral functions from correlation functions,
the corresponding formalism and procedure are presented in Sec.~\ref{subsec:formalism_SpM}.

\subsection{
Brief review of the sparse modeling method
\label{subsec:review_SpM}
}
Extracting spectral functions by using SpM has been proposed in condensed matter physics~\cite{Otsuki.PhysRevE.95.061302, Shinaoka.PhysRevB.96.035147}.
Following Ref.~\cite{Ohtsuki_JPSJ.89.012001}
which provides a more detailed review,
we give a brief review on formulation of the SpM method in the IR basis.
\\
\indent
Euclidean formulation of lattice QCD
cannot directly determine the meson spectral function $\rho(\omega)$ at the frequency $\omega$;
instead, one can measure the meson correlation function $G(\tau)$ at the Euclidean time $\tau$.
$G(\tau)$ and $\rho(\omega)$ can be written in Lehmann representation
in the following form
\begin{equation}
  G(\tau)=\int^{\infty}_{0}d\omega K(\tau,\omega)\rho(\omega),
  \label{eq:spf_Lehmann_rep_0}
\end{equation}
by using the Kubo-Martin-Schwinger relation
and the periodic boundary condition of $G(\tau)$.
Here,
$K(\omega,\tau)$ is the integration kernel represented
in the following form
\begin{equation}
  \displaystyle K(\tau,\omega)
  =\frac{\cosh\left[
      \omega\left(
      \tau-\frac{1}{2T}
      \right)
      \right]}{\sinh\left(
    \frac{\omega}{2T}
    \right)},
  \label{eq:1_kernel}
\end{equation}
in the Euclidean time range $0\le\tau\le 1/T$ with temperature $T$.
Note that
the kernel $K(\tau,\omega)$ is symmetric around $\tau=1/(2T)$
and the spectral function $\rho(\omega)$
is an odd function of the frequency $\omega$.
This means that $\rho(\omega)$ holds
\begin{equation}
  \frac{\rho(\omega)}{\omega}\ge 0.
\end{equation}
Equations~\eqref{eq:spf_Lehmann_rep_0} and~\eqref{eq:1_kernel}
are used as clues to determine $\rho(\omega)$
using $G(\tau)$ as the input data.
\\
\indent
In numerical calculations
the Euclidean time $\tau$ and the frequency $\omega$ must be discretized.
The length in the Euclidean-time direction is limited
by the lattice size of the lattice simulations,
and we introduce a cutoff $\omega_{\mathrm{max}}$ for the infinite integral over $\omega$.
Then the variable $\tau$ and $\omega$ are discretized
into $M$ and $N$ slices, respectively.
Defining $G_{i}\equiv G(\tau_{i})$ and $\rho_{i}\equiv\rho(\omega_{i})\Delta\omega$
where $\Delta\omega$ is the interval of $\omega$ instead of $d\omega$,
we can denote Eq.~\eqref{eq:spf_Lehmann_rep_0} using a matrix-vector representation
\begin{equation}
  \vec{G}=K\vec{\rho}.
  \label{eq:linear_eq}
\end{equation}
Here,
$K$ is an $M\times N$ matrix defined by $K_{ij}\equiv K(\tau_{i},\omega_{j})$.
\\
\indent
To find $\vec{\rho}$,
we consider the square error
\begin{equation}
    \chi^{2}(\vec{\rho} \mid\vec{G})
    =\frac{1}{2}||\vec{G}-K\vec{\rho}||^{2}_{2},
    \label{eq:chi2}
\end{equation}
and minimize it.
Here,
the argument on the left-hand side of the bar ($\vec{\rho}$)
indicates a quantity to be varied,
that on the right-hand side of the bar ($\vec{G}$)
indicates a fixed quantity
and $||\cdot ||_{2}$ stands for the $L_{2}$ norm
defined by
\begin{equation}
    ||\vec{\rho}\ ||_{2}\equiv\Bigl(
        \sum_{j}\rho_{j}^{2}
    \Bigr)^{1/2}.
\end{equation}
Since we now assuming that $N>M$,
the number of equation~\eqref{eq:linear_eq}
is insufficient for $\vec{\rho}$ to be determined uniquely.
We could solve the equation if $N=M$,
while $\vec{G}$ has a noise,
and it is difficult to obtain the stable solution of $\vec{\rho}$.
\\
\indent
To address these difficulties
we take an efficient basis that yields a sparse solution.
Here,
``sparse'' is such that most components of the solution are zero.
If we can find the position of zeros in the components
and remove these zeros from the set of equations,
the number of components will be reduced,
and then equations are solvable if $N=M$.
\\
\indent
For the realisation of this situation,
we firstly decompose the kernel $K$ using the singular value decomposition (SVD) as
\begin{equation}
    K=USV^{\mathrm{t}},
    \label{eq:SVD}
\end{equation}
where the superscript t indicates transpose,
$S$ is an $M\times N$ matrix composed of singular values
$s_{l}\; (l=1,2,\cdots,M)$ at the diagonal,
and $U$ and $V$ are $M\times M$ and $N\times N$ orthogonal matrices,
respectively.
Here, the sequence $\{s_{l}\}$ is sorted by magnitude in descending order.
Then we transform the vectors $\vec{G}$ and $\vec{\rho}$
using the orthogonal matrices $U$ and $V$ as
\begin{equation}
  \vec{G}^{\prime}\equiv U^{\mathrm{t}}\vec{G},
   \quad
  \vec{\rho}^{\;\prime}\equiv V^{\mathrm{t}}\vec{\rho},
  \label{eq:X_Grho}
\end{equation}
respectively.
This new basis is called intermediate representation (IR) basis.
Rewriting the square error in Eq.~\eqref{eq:chi2} with the new vectors,
$\vec{G}^{\prime}$ and $\vec{\rho}^{\;\prime}$,
in the IR basis,
we obtain 
\begin{align}
    \chi^{2}(\vec{\rho}^{\;\prime}\mid\vec{G}^{\prime})
    &=\frac{1}{2}||\vec{G}^{\prime}-S\vec{\rho}^{\;\prime}||^{2}_{2}\\
    &=\frac{1}{2}\sum_{l}(G^{\prime}_{l}-s_{l}\rho^{\prime}_{l})^{2}.
    \label{eq:chi2_IR}
\end{align}
Thus minimizing $\chi^{2}(\vec{\rho}^{\;\prime}\mid\vec{G}^{\prime})$
is to find a solution that satisfies the equation 
\begin{equation}
    G^{\prime}_{l}=s_{l}\rho^{\prime}_{l}.
    \label{eq:G'=srho'}
\end{equation}
Note that the singular values $s_{l}$ for the type of kernel
described by Eq.~\eqref{eq:1_kernel} decrease exponentially.
\\
\indent
We focus on the property of $s_{l}$.
$G^{\prime}_{l}$ can decay much faster than $\rho^{\prime}_{l}$
due to the behavior of $s_{l}$,
which relies on the nature of the kernel $K$.
Thus,
the fast decay of $G^{\prime}_{l}$ does not depend on a simulation detail.
It indicates that
$G^{\prime}_{l}$ does not include information about $\rho^{\prime}_{l}$
at large $l$.
In other words,
elements of $G^{\prime}_{l}$, $s_{l}$ and $\rho^{\prime}_{l}$
can be removed at large $l$.
It is called ``dimensionality reduction'' in Ref.~\cite{Ohtsuki_JPSJ.89.012001}.
This reduction is a good solution for the inverse problem we address,
which involves correlation functions with noise as input.
Equation~\eqref{eq:G'=srho'} shows
that $\rho(\omega)$ and $G(\tau)$ are directly connected to each other in the IR basis,
and $\rho(\omega)$ could be naively evaluated by using the relation
\begin{equation}
    \rho^{\prime}_{l}=G^{\prime}_{l}/s_{l}.
    \label{eq:rho'=G'/s}
\end{equation}
This evaluation, however,
does not work well due to noise of $G(\tau)$.
Even if the noise in $G(\tau)$ is small,
its influence is amplified at large $l$,
as Eq.~\eqref{eq:rho'=G'/s} shows.
On the other hand,
the influences of noise can be mitigated by the reduction.
\\
\indent
In actual calculations,
thresholds $s_{\mathrm{cut}}$ are set for the obtained singular values $s_{l}$,
taking into account the properties described above.
We can discard such a component corresponding to $l>l_{\mathrm{cut}}$
below this threshold,
$s_{l}<s_{\mathrm{cut}}$.
This truncation is necessary to obtain a stable solution against the noise in $G(\tau)$.
\\
\indent
After dimensionality reduction,
imposing sparseness on $\vec{\rho}^{\;\prime}$ to obtain a stable solution against noise,
we introduce the $L_{1}$ regularization term to $\chi^{2}(\vec{\rho}^{\;\prime}\mid\vec{G}^{\prime})$ as
\begin{equation}
    F(\vec{\rho}^{\;\prime}\mid\vec{G}^{\prime},\lambda)
    \equiv\chi^{2}(\vec{\rho}^{\;\prime}\mid\vec{G}^{\prime})
    +\lambda||\vec{\rho}^{\;\prime}||_{1},
    \label{eq:cost_function_F}
\end{equation}
where $\lambda$ is a positive hyperparameter
which controls the contribution of $L_{1}$ regularization
relative to $\chi^{2}(\vec{\rho}^{\;\prime}\mid\vec{G}^{\prime})$,
and $||\cdot||_{1}$ stands for the $L_{1}$ norm defined by
\begin{equation}
    ||\vec{\rho}^{\;\prime}||_{1}
    \equiv\sum_{l}|\rho^{\prime}_{l}|.
\end{equation}
The $L_{1}$ regularization makes the solution sparse and remove irrelevant bases.
Thus the parameter $\lambda$ regulates the degree of sparseness.
The optimal value of $\lambda$ can be automatically determined 
as will be mentioned in Sec.~\ref{subsec:formalism_SpM}.
This minimization problem,
expressed by Eq.~\eqref{eq:cost_function_F},
is known as the least absolute shrinkage and selection operator (LASSO).
Given that the LASSO is a convex optimization problem,
the global minimum can be achieved
regardless of initial conditions~\cite{Boyd_Vandenberghe_cvx_opt}.
\\
\indent
The solution will hold two additional constraints.
One is the non-negativity condition of the spectral function,
\begin{equation}
    \rho_{j}\ge 0.
    \label{eq:rho_positivity}
\end{equation}
This condition is imposed in this study.
The other is the sum rule,
\begin{equation}
    \sum_{j}\rho_{j}=1,
    \label{eq:rho_sum_rule}
\end{equation}
while this condition is not imposed in this study.
With these additional constraints,
the minimization problem in Eq.~\eqref{eq:cost_function_F}
can be solved 
using the alternating direction method of multipliers (ADMM) algorithm~\cite{Boyd.MAL-016}.
For the detail of the algorithm,
see appendix~\ref{sec:ADMM}.

\subsection{
Consideration of covariance matrices for sparse modeling
\label{subsec:consider_C_for_SpM}
}
In practical simulations,
we obtain the Monte Carlo data $\vec{G}$ with noise
which is the source of the statistical error,
and $\vec{G}$ has correlation between different Euclidean times.
Thus the covariance of $\vec{G}$ should be taken into account.
In this case
Ref.~\cite{Ohtsuki_JPSJ.89.012001} explains
that the square error in Eq.~\eqref{eq:chi2} can be extent 
to a form,
\begin{equation}
    \chi^{2}(\vec{\rho} \mid \vec{G},C)
    =\frac{1}{2}(\vec{G}-K\vec{\rho})^{\mathrm{t}}
    C^{-1}(\vec{G}-K\vec{\rho}),
    \label{eq:chi2wC}
\end{equation}
where $C$ represents the covariance matrix defined by
\begin{align}
  C_{ij}=&\frac{1}{N_{\mathrm{conf}}(N_{\mathrm{conf}}-1)}
  \nonumber\\
  &\times\sum^{N_{\mathrm{conf}}}_{n=1}
  \left(G_{i}-G^{(n)}_{i}\right)\left(G_{j}-G^{(n)}_{j}\right),
  \label{eq:Cij}\\
  G_{i}=&\frac{1}{N_{\mathrm{conf}}}
  \sum^{N_{\mathrm{conf}}}_{n=1}G^{(n)}_{i},
\end{align}
where $N_{\mathrm{conf}}$ is the total number of gauge configurations
and $G_i^{(n)}$ is the value of the correlation function
measured on the $n$-th gauge configuration.
\\
\indent
Since the covariance matrix $C$ is a symmetric matrix,
its inverse is also symmetric
and can be factorized using the Cholesky decomposition as
\begin{equation}
    C^{-1}=W^{\mathrm{t}}W,
    \label{eq:CD}
\end{equation}
where $W$ is the upper triangular matrix.
Using Eq.~\eqref{eq:CD}
and defining
\begin{equation}
    \vec{G}_{W}\equiv W\vec{G},
    \quad
    K_{W}\equiv WK,
    \label{eq:GW=WG,KW=WK}
\end{equation}
we can rewrite Eq.~\eqref{eq:chi2wC} as
\begin{equation}
    \chi^{2}(\vec{\rho} \mid \vec{G}_{W})
    =\frac{1}{2}||\vec{G}_{W}-K_{W}\vec{\rho}||^{2}_{2}.
    \label{eq:chi2wW}
\end{equation}
Now that we have the squared error such as Eq.~\eqref{eq:chi2},
by decomposing $K_{W}$ using SVD as
\begin{equation}
    K_{W}=U_{W}S_{W}V_{W}^{\mathrm{t}},
\end{equation}
where $S_{W}$ is an $M\times N$ matrix composed singular values,
and $U_{W}$ and $V_{W}$ are $M\times M$ and $N\times N$ orthogonal matrices,
respectively,
and defining new vectors in the IR basis as
\begin{equation}
    \vec{G}^{\prime}_{W}\equiv U_{W}^{\mathrm{t}}\vec{G}_{W},
    \quad
    \vec{\rho}^{\;\prime}_{W}\equiv V_{W}^{\mathrm{t}}\vec{\rho},
\end{equation}
we can express the squared error
\begin{equation}
    \chi^{2}(\vec{\rho}^{\;\prime}_{W}\mid\vec{G}^{\prime}_{W})
    =\frac{1}{2}||\vec{G}^{\prime}_{W}-S_{W}\vec{\rho}^{\;\prime}_{W}||^{2}_{2},
    \label{eq:chi2wW_IR}
\end{equation}
in the IR basis.
The optimization problem to find a sparse solution of $\vec{\rho}^{\;\prime}_{W}$
can be solved using 
\begin{equation}
    F(\vec{\rho}^{\;\prime}_{W}\mid\vec{G}^{\prime}_{W},\lambda)
    \equiv\chi^{2}(\vec{\rho}^{\;\prime}_{W}\mid\vec{G}^{\prime}_{W})
    +\lambda||\vec{\rho}^{\;\prime}_{W}||_{1},
    \label{eq:cost_function_FW}
\end{equation}
with two additional constraints in Eqs.~\eqref{eq:rho_positivity} and~\eqref{eq:rho_sum_rule}.
\\
\indent
In order to use dimensionality reduction,
it is necessary that the components of $S_{W}$ decrease exponentially.
We make sure that in the Appendix~\ref{sec:SW_of_KW}.

\subsection{
Formalism of sparse modeling for extraction of spectral functions
\label{subsec:formalism_SpM}
}
Here we present the SpM formalism
to extract spectral functions
from correlation functions obtained by lattice calculations.
\\
\indent
The starting point of the formalism is
Eq.~\eqref{eq:spf_Lehmann_rep_0}.
According to the detailed formulation described in Appendix~\ref{sec:formalism_ap},
equation~\eqref{eq:spf_Lehmann_rep_0} can be written as
\begin{align}
    \hat{G}(\hat{\tau}_{i})
    &=\sum^{N_{\omega}-1}_{j=0}
    K^{(0)}(\hat{\tau}_{i},\hat{\omega}_{j})
    \tilde{\rho}^{(0)}(\hat{\omega}_{j}),
    \label{eq:G_hat_tau_hat}\\
    K^{(0)}(\hat{\tau}_{i},\hat{\omega}_{j})
    &=\sqrt{\Delta\omega^{\prime}}\hat{\omega}_{\mathrm{max}}
    \frac{\cosh\left[
      \hat{\omega}_{j}\left(
      \hat{\tau}_{i}-N_{\tau}/2
      \right)
      \right]}{\cosh\left[
      \hat{\omega}_{j}\left(
      \hat{\tau}_{\mathrm{r}}-N_{\tau}/2
      \right)
    \right]},
    \label{eq:K_tau_hat_omega_hat}\\
    \tilde{\rho}^{(0)}(\hat{\omega}_{j})
    &=\sqrt{\Delta\omega^{\prime}}
    \frac{\cosh\left[
      \hat{\omega}_{j}\left(
      \hat{\tau}_{\mathrm{r}}-N_{\tau}/2
      \right)
      \right]}{\sinh\left(
    \hat{\omega}_{j}N_{\tau}/2
    \right)}
    \hat{\rho}(\hat{\omega}_{j}),
    \label{eq:rho_tilde_omega_hat}
\end{align}
where hatted variables are dimensionless quantities,
$N_{\omega}$ is the number of points in $\omega$-space,
$\Delta\omega^{\prime}\equiv 1/(N_{\omega}-1)$,
$\hat{\omega}_{\mathrm{max}}$ is a cutoff for $\hat{\omega}$,
$N_{\tau}$ is the extent of Euclidean times
and $\hat{\tau}_{\mathrm{r}}$ is the reference Euclidean time.
In this study,
we extract $\tilde{\rho}^{(0)}(\hat{\omega}_{j})$ directly
from correlation functions using ADMM algorithm for $K^{(0)}(\hat{\tau}_{i},\hat{\omega}_{j})$.
\\
\indent
Note that
$\tilde{\rho}^{(0)}(\hat{\omega}_{j})$ in Eq.~\eqref{eq:rho_tilde_omega_hat}
is proportional to $\hat{\rho}(\hat{\omega}_{j})/\hat{\omega}_{j}$
in the limit $\hat{\omega}\to 0$.
One method of investigating the transport properties
at temperatures above $T_{\mathrm{c}}$ is to examine the transport peak of the spectral function.
Theoretically,
the transport peak has a finite value
in $\hat{\rho}(\hat{\omega}_{j})/\hat{\omega}_{j}$ at $\hat{\omega} \to 0$.
Thus,
it is favorable to find the equivalent of $\hat{\rho}(\hat{\omega}_{j})/\hat{\omega}_{j}$ directly
using ADMM algorithm from the outset,
and $\tilde{\rho}^{(0)}(\hat{\omega}_{j})$ in Eq.~\eqref{eq:rho_tilde_omega_hat}
is of that form.
\\
\indent
At temperatures below $T_{\mathrm{c}}$,
$\hat{\rho}(\hat{\omega}_{j})$ is expected to be asymptotically proportional to $\hat{\omega}^{2}$ in high frequency region.
Therefore,
the resonance peak is usually checked by $\hat{\rho}(\hat{\omega}_{j})/\hat{\omega}_{j}^{2}$.
Thus,
it is favorable to find the equivalent
of $\hat{\rho}(\hat{\omega}_{j})/\hat{\omega}_{j}^{2}$ directly
using ADMM algorithm.
In addition,
a transport peak for the spectral function is expected to not appear at $\omega\to 0$
at temperatures below $T_{\mathrm{c}}$.
Taken these features into consideration,
equations~\eqref{eq:G_hat_tau_hat}-\eqref{eq:rho_tilde_omega_hat} can be rewritten as
\begin{align}
    \hat{G}(\hat{\tau}_{i})
    &=\sum^{N_{\omega}-1}_{j=0}
    K^{(1)}(\hat{\tau}_{i},\hat{\omega}_{j})
    \tilde{\rho}^{(1)}(\hat{\omega}_{j}),
    \label{eq:G_hat_tau_hat11}\\
    K^{(1)}(\hat{\tau}_{i},\hat{\omega}_{j})
    &=\sqrt{\Delta\omega^{\prime}}\hat{\omega}_{j}\hat{\omega}_{\mathrm{max}}
    \frac{\cosh\left[
      \hat{\omega}_{j}\left(
      \hat{\tau}_{i}-N_{\tau}/2
      \right)
      \right]}{\cosh\left[
      \hat{\omega}_{j}\left(
      \hat{\tau}_{\mathrm{r}}-N_{\tau}/2
      \right)
    \right]},
    \label{eq:K1_tau_hat_omega_hat}\\
    \tilde{\rho}^{(1)}(\hat{\omega}_{j})
    &=\sqrt{\Delta\omega^{\prime}}
    \frac{\cosh\left[
      \hat{\omega}_{j}\left(
      \hat{\tau}_{\mathrm{r}}-N_{\tau}/2
      \right)
      \right]}{\sinh\left(
    \hat{\omega}_{j}N_{\tau}/2
    \right)}
    \frac{\hat{\rho}(\hat{\omega}_{j})}{\hat{\omega}_{j}}.
    \label{eq:rho1_tilde_omega_hat}
\end{align}
$K^{(1)}(\hat{\tau}_{i},\hat{\omega}_{j})$ incorporates an additional $\hat{\omega}_{j}$
with respect to $K^{(0)}(\hat{\tau}_{i},\hat{\omega}_{j})$,
and $\tilde{\rho}^{(0)}(\hat{\omega}_{j})$ is divided by $\hat{\omega}_{j}$
to yield $\tilde{\rho}^{(1)}(\hat{\omega}_{j})$.
\\
\indent
Note that $\tilde{\rho}^{(1)}(0)=0$ is obtained in SpM analysis.
Since $K^{(1)}(\hat{\tau}_{i},0)=0$,
$\tilde{\rho}^{(1)}(0)$ does not contribute to the correlation function.
Conversely, the values of $\tilde{\rho}^{(1)}(0)$ cannot be constrained from the given data.
On the other hand,
attempting to set $\tilde{\rho}^{(1)}(0)$ to a non-zero value inevitably increases the error function due to the $L_1$ regularization term.
Therefore, the most favorable solution is to set $\tilde{\rho}^{(1)}(0)$ to zero.
Furthermore, due to dimensionality reduction,
SpM uses only the minimum necessary modes,
and high-frequency modes sensitive to noise are prioritize for removal during this process. 
Thus, sharp structures requiring many high-frequency modes are less likely to appear,
and a smooth spectral function can be obtained.
Consequently, $\tilde{\rho}^{(1)}(\hat{\omega}_{j})$ smoothly approaches zero near $\hat{\omega}=0$.
Based on these considerations,
in this study,
$\tilde{\rho}^{(1)}(\hat{\omega}_{j})$ is also extracted directly
from correlation functions at temperature below $T_{\mathrm{c}}$
using ADMM algorithm for $K^{(1)}(\hat{\tau}_{i},\hat{\omega}_{j})$.
However, it is clear that $K^{(1)}(\hat{\tau}_{i},\hat{\omega}_{j})$ should not be employed in analyses at temperature above $T_{\mathrm{c}}$,
where a transport peak is expected to exist.
\\
\indent
The covariance matrix $C$ is considered in our SpM analyses.
To account for the covariance in the actual calculation,
rewriting $K_{W}$ defined in Eq.~\eqref{eq:GW=WG,KW=WK},
we define $K^{(n)}_{W}(\hat{\tau}_{k},\hat{\omega}_{j})$ as
\begin{equation}
    K^{(n)}_{W}(\hat{\tau}_{k},\hat{\omega}_{j})
    \equiv\sum_{i=0}^{N_{\tau}-1}
    W(\hat{\tau}_{k},\hat{\tau}_{i})K^{(n)}(\hat{\tau}_{i},\hat{\omega}_{j}),
    \label{eq:kernel_W}
\end{equation}
where $n=0,1$
and the matrix $W(\hat{\tau}_{k},\hat{\tau}_{i})$
is obtained from the Cholesky decomposition of $C^{-1}$ as
\begin{equation}
    C^{-1}(\hat{\tau}_{k},\hat{\tau}_{l})
    =\sum_{m=0}^{N_{\tau}-1}W(\hat{\tau}_{m},\hat{\tau}_{k})
    W(\hat{\tau}_{m},\hat{\tau}_{l}),
\end{equation}
which is rewritten from Eq.~\eqref{eq:CD}.

\subsection{
Procedure of sparse modeling for extraction of spectral functions
\label{subsec:procedure_SpM}
}
Now, we summarize the SpM procedure in this study.
The following procedure assumes that
the correlation functions $\hat{G}(\hat{\tau}_{i})$ have been obtained by lattice calculation,
and superscripts $(n)$ introduced in Sec.~\ref{subsec:formalism_SpM} are omitted here
for simplicity.
\begin{enumerate}
  \item Calculate the covariance matrix $C$ from the correlation functions $\hat{G}(\hat{\tau}_{i})$,
  carry out the Cholesky decomposition of the inverse of $C$
  and obtain the matrix $W$.
  \item Construct the discretized kernel $K_{W}$ using the kernel $K$ and $W$.
  \item Perform SVD of the kernel $K_{W}$.
  \item Transform the basis of $\hat{G}(\hat{\tau}_{i})$
  and $\tilde{\rho}(\hat{\omega}_{j})$
  by $U^{\mathrm{t}}_{W}$ and $V^{\mathrm{t}}_{W}$
  to obtain $G^{\prime}_{W}(\hat{\tau}_{i})$
  and $\rho^{\prime}_{W}(\hat{\omega}_{j})$ in IR basis.
  \item Choose up to $L$-th largest singular values
  satisfied with the condition $s_{l}/s_{1}\ge 10^{-15}$,
  where $s_{l}$ is the $l$-th largest singular value,
  and drop the components of $G^{\prime}_{W}(\hat{\tau}_{i})$
  and $\rho^{\prime}_{W}(\hat{\omega}_{j})$
  corresponding to the other small singular values,
  which reduces the size of $U_{W}$, $V_{W}$ and $S_{W}$ to $N_{\tau}\times L$,
  $N_{\omega}\times L$ and $L\times L$,
  respectively.
  \item Construct the cost function
  $F(\rho^{\prime}_{W}(\hat{\omega}_{j})\mid G^{\prime}_{W}(\hat{\tau}_{i}),\lambda)$
  from the square error $\chi^2(\rho^{\prime}_{W}(\hat{\omega}_{j})\mid G^{\prime}_{W}(\hat{\tau}_{i}))$
  and the $L_{1}$ regularization term.
  \item Estimate the optimal value of $\lambda$,
  $\lambda_\mathrm{opt}$,
  in the same way as in the previous study~\cite{itou2020sparse},
  which is explained in detail in Appendix~\ref{sec:lambda_opt}.
  \item Find the most likely spectral function $\rho^{\prime}_{W}(\hat{\omega}_{j})$
  in IR basis by minimizing the cost function
  $F(\rho^{\prime}_{W}(\hat{\omega}_{j})\mid G^{\prime}_{W}(\hat{\tau}_{i}),\lambda_{\mathrm{opt}})$
  using the ADMM algorithm,
  transform the basis of $\rho^{\prime}_{W}(\hat{\omega}_{j})$ by $V_{W}$
  to obtain $\tilde{\rho}(\hat{\omega}_{j})$
  and calculate $\hat{\rho}(\hat{\omega}_{j})$.
\end{enumerate}

\section{
Mock data Analysis
\label{sec:mock_data_tests}
}
In this section we show the results of the mock-data tests using SpM.
As mentioned above,
SpM has been applied to correlation functions 
computed on actual lattice QCD data only in a few cases, 
and meson spectral functions at finite $T$ have not yet been computed by using this method.
Therefore, 
before we analyze the actual lattice QCD data of meson correlation functions by using SpM,
we first test the method with mock data that mimic possible charmonium spectral functions.
\\
\indent
Since we are interested in the applicability of this method to meson correlation functions,
we will prepare spectral functions that reproduce correlation functions
close to those computed for lattice QCD.
This spectral function is shown in Sec.~\ref{subsec:mock_spf}.
Another aspect is applicability with respect to the number of data points and the quality of the data.
The results of these investigations are discussed in Sec.~\ref{subsec:dependence_Nt_epsilon}.
A further aspect examined in this work is the reproducibility of the anticipated peak structure.
In Sec.~\ref{subsec:validation_spf},
this issue is addressed by means of analyses of mock data generated from free spectral functions without peaks to see how reliably SpM reconstructs the spectral feature.

\subsection{
Mock spectral functions and correlation functions
\label{subsec:mock_spf}
}
Following Ref.~\cite{Ding2018.PhysRevD.97.094503}
we consider the input spectral functions
at temperatures below and above $T_{\mathrm{c}}$:
\begin{itemize}
    \item For $T<T_{\mathrm{c}}$
    \begin{align}
        \hat{\rho}_{\mathrm{below}}(\hat{\omega})
        =&\tilde{\Theta}(\hat{\omega},\hat{\omega}_{1},\Delta_{1})
        (1-\tilde{\Theta}(\hat{\omega},\hat{\omega}_{2},\Delta_{2}))
        \hat{\rho}_{\mathrm{res}}
        \nonumber\\
        &+\tilde{\Theta}(\hat{\omega},\hat{\omega}_{3},\Delta_{3})
        \hat{\rho}_{\mathrm{Wilson}}
        \label{eq:rho_below_fw}
    \end{align}
    \item For $T>T_{\mathrm{c}}$
    \begin{align}
        \hat{\rho}_{\mathrm{above}}(\hat{\omega})
        =&\hat{\rho}_{\mathrm{trans}}
        \nonumber\\
        &+\tilde{\Theta}(\hat{\omega},\hat{\omega}_{4},\Delta_{4})
        (1-\tilde{\Theta}(\hat{\omega},\hat{\omega}_{5},\Delta_{5}))
        \hat{\rho}_{\mathrm{res}}
        \nonumber\\
        &+\tilde{\Theta}(\hat{\omega},\hat{\omega}_{6},\Delta_{6})
        \hat{\rho}_{\mathrm{Wilson}}
        .
        \label{eq:rho_above_fw}
    \end{align}
\end{itemize}
Here $\hat{\rho}_{\mathrm{res}}$ denotes a resonance peak,
$\hat{\rho}_{\mathrm{trans}}$ denotes a transport peak
and $\hat{\rho}_{\mathrm{Wilson}}$ denotes a free Wilson spectral function~\cite{PhysRevD.68.014504,AARTS200593} 
which takes the lattice cutoff effect into account.
In ref.~\cite{Ding2018.PhysRevD.97.094503}
a free continuum spectral function is used in the $\hat{\rho}_{\mathrm{below}}$,
but we use the free Wilson spectral function
to account for the lattice cutoff effects even in the case of $T<T_{\mathrm{c}}$.
\\
\indent
The spectral functions are needed to be smooth, 
and the $\hat{\rho}_{\mathrm{res}}$, $\hat{\rho}_{\mathrm{trans}}$ and $\hat{\rho}_{\mathrm{Wilson}}$
that make up $\hat{\rho}_{\mathrm{below}}$ and $\hat{\rho}_{\mathrm{above}}$ 
are connected by a modified $\Theta$ function,
which is given by
\begin{align}
    \tilde{\Theta}(\hat{\omega},\hat{\omega}_{i},\Delta_{i})
   =\left(
        1+\exp\left(
            \frac{\hat{\omega}_{i}^{2}-\hat{\omega}^{2}}{\hat{\omega}\Delta_{i}}
        \right)
    \right)^{-1}.
\end{align}
The $\hat{\rho}_{\mathrm{res}}$, $\hat{\rho}_{\mathrm{trans}}$
and $\hat{\rho}_{\mathrm{Wilson}}$ are listed below:
\begin{itemize}
    \item Transport peak
    \begin{equation}
        \hat{\rho}_{\mathrm{trans}}
        =c_{\mathrm{trans}}\frac{\hat{\omega}\eta}{\hat{\omega}^{2}+\eta^{2}}.
    \end{equation}
    \item Resonance peak
    \begin{equation}
        \hat{\rho}_{\mathrm{res}}
        =c_{\mathrm{res}}\frac{\Gamma M\hat{\omega}^{2}}{(\hat{\omega}^{2}-M^{2})^{2}+M^{2}\Gamma^{2}}.
    \end{equation}
    \item Free Wilson spectral function
    \begin{align}
        \hat{\rho}_{\mathrm{Wilson}}
        =&c_{\mathrm{Wilson}}\frac{4\pi N_{c}}{N_{\sigma}^{3}}
        \sum_{\vec{k}}\sinh\left(
            \frac{\hat{\omega}}{2\hat{T}}
        \right)
        \nonumber\\
        &\times\left[
            b^{(1)}-b^{(2)}
            \frac{\sum_{i=1}^{3}\sin^{2}k_{i}}{\sinh^{2}E_{\vec{k}}(m)}
        \right]
        \nonumber\\
        &\times\frac{\delta(\hat{\omega}-2E_{\vec{k}}(m))}
        {2(1+M_{\vec{k}}(m))^{2}\cosh^{2}\left(
            \frac{E_{\vec{k}}(m)}{2\hat{T}}
        \right)},
    \end{align}
    where
    \begin{align}
        \cosh E_{\vec{k}}(m)
        &=1+\frac{K_{\vec{k}}^{2}+M_{\vec{k}}^{2}(m)}{2(1+M_{\vec{k}}(m))},\\
        K_{\vec{k}}&=\sum_{i=1}^{3}\gamma_{i}\sin k_{i},\\
        M_{\vec{k}}(m)&=\sum_{i=1}^{3}(1-\cos k_{i})+m.
    \end{align}
\end{itemize}
\indent
The values of the parameters used in the above spectral function
are summarized in table~\ref{tab:params_mock_spf}.
These parameters mimic the physical situation at lattice spacing $a^{-1}=20$ GeV,
reflecting the fact that
the position of the resonance peak is approximately $J/\psi$ meson mass ($\sim 3.1$ GeV) for $T < T_{\mathrm{c}}$,
and that a transport peak appears and the resonance peak becomes broader for $T > T_{\mathrm{c}}$.
In the free Wilson spectral function,
the quark mass is set to be about 1.5 GeV,
and the threshold of this spectral function can be adjusted
by using the modified $\Theta$ function in the bound state region.
The values of the parameters used in the $\Theta$ function
are summarized in table~\ref{tab:params_modified_theta}.
\renewcommand{\arraystretch}{1.2}
\begin{table*}[htbp]
    \centering
    \begin{tabular}{r|c}
    \hline
    \hline
        Spectral function & Parameters \\
    \hline
        $\hat{\rho}_{\mathrm{res}}$ for $\hat{\rho}_{\mathrm{below}}$ & $c_{\mathrm{res}}=0.08/7$, $\Gamma=0.05$, $M=0.155$ \\
        $\hat{\rho}_{\mathrm{Wilson}}$ for $\hat{\rho}_{\mathrm{below}}$ & $c_{\mathrm{Wilson}}=0.5$, $b^{(1)}=2$, $b^{(2)}=1$, $m=0.073$, $N_{c}=3$, $N_{\sigma}=4096$\\
        $\hat{\rho}_{\mathrm{trans}}$ for $\hat{\rho}_{\mathrm{above}}$ & $c_{\mathrm{trans}}=5\times 10^{-5}$, $\eta=0.006$\\
        $\hat{\rho}_{\mathrm{res}}$ for $\hat{\rho}_{\mathrm{above}}$ & $c_{\mathrm{res}}=0.06$, $\Gamma=0.15$, $M=0.225$ \\
        $\hat{\rho}_{\mathrm{Wilson}}$ for $\hat{\rho}_{\mathrm{above}}$ & $c_{\mathrm{Wilson}}=1$, $b^{(1)}=3$, $b^{(2)}=1$, $m=0.073$, $N_{c}=3$, $N_{\sigma}=4096$\\
    \hline
    \hline
    \end{tabular}
    \caption{
    Parameters for the mock spectral functions.
    }
    \label{tab:params_mock_spf}
\end{table*}

\begin{table}[htb]
    \centering
    \begin{tabular}{|c|c|}
    \hline
        $\hat{\omega}_{1}=0.145$ & $\Delta_{1}=0.01$ \\
        $\hat{\omega}_{2}=0.155$ & $\Delta_{2}=0.05$ \\
        $\hat{\omega}_{3}=0.225$ & $\Delta_{3}=0.05$ \\
        $\hat{\omega}_{4}=0.225$ & $\Delta_{4}=0.15$ \\
        $\hat{\omega}_{5}=0.225$ & $\Delta_{5}=0.15$ \\
        $\hat{\omega}_{6}=0.350$ & $\Delta_{6}=0.20$ \\
        \hline
    \end{tabular}
    \caption{
    Parameters for $\tilde{\Theta}$ functions.
    }
    \label{tab:params_modified_theta}
\end{table}
\renewcommand{\arraystretch}{1}

\indent
The mock correlation function and the covariance matrix are obtained as follows.
First,
the correlation function is calculated
by integrating the spectral function and the kernel $K$
(see Eq.~\eqref{eq:G_hat_tau_hat} or Eq.~\eqref{eq:G_hat_tau_hat11}). 
Then, Gaussian random numbers with standard deviation
$\sigma_{i}=\varepsilon\cdot G(\tau_{i})\cdot\tau_{i}$ are created at each Euclidean time.
They correspond to the error
with respect to the central value of the correlation function $G(\tau_{i})$.
In this study,
we have 300 errors.
These are added to the correlation function,
and then 300 correlation functions $G^{(n)}(\tau_{i})$ with errors,
i.e., $N_{\mathrm{conf}}=300$, are obtained.
Using these correlation functions,
we can calculate the covariance matrix $C_{ij}$
by computing Eq.~\eqref{eq:Cij}.
In our mock-data tests,
$\hat{\tau}_{\mathrm{r}}$ is set to 1,
and also $\hat{\tau}_{\mathrm{min}}$ and $\hat{\tau}_{\mathrm{max}}$ are set
to 1 and $N_{\tau}/2$, respectively.

\subsection{
Dependence on $N_{\tau}$ and noise level $\varepsilon$
\label{subsec:dependence_Nt_epsilon}
}
\begin{figure*}[tb]
  \centering
  \begin{minipage}{.49\textwidth}
    \includegraphics[width=1.0\linewidth]{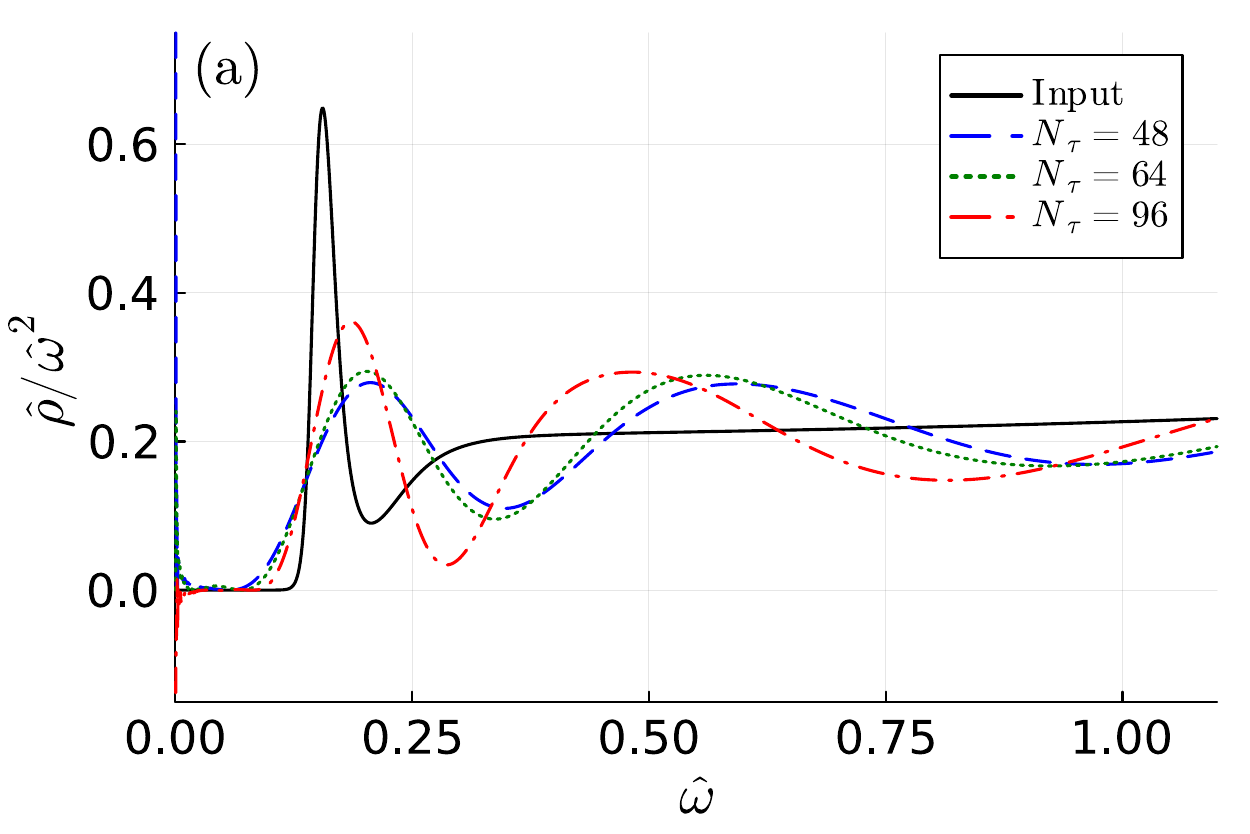}
  \end{minipage}
  \begin{minipage}{.49\textwidth}
    \includegraphics[width=1.0\linewidth]{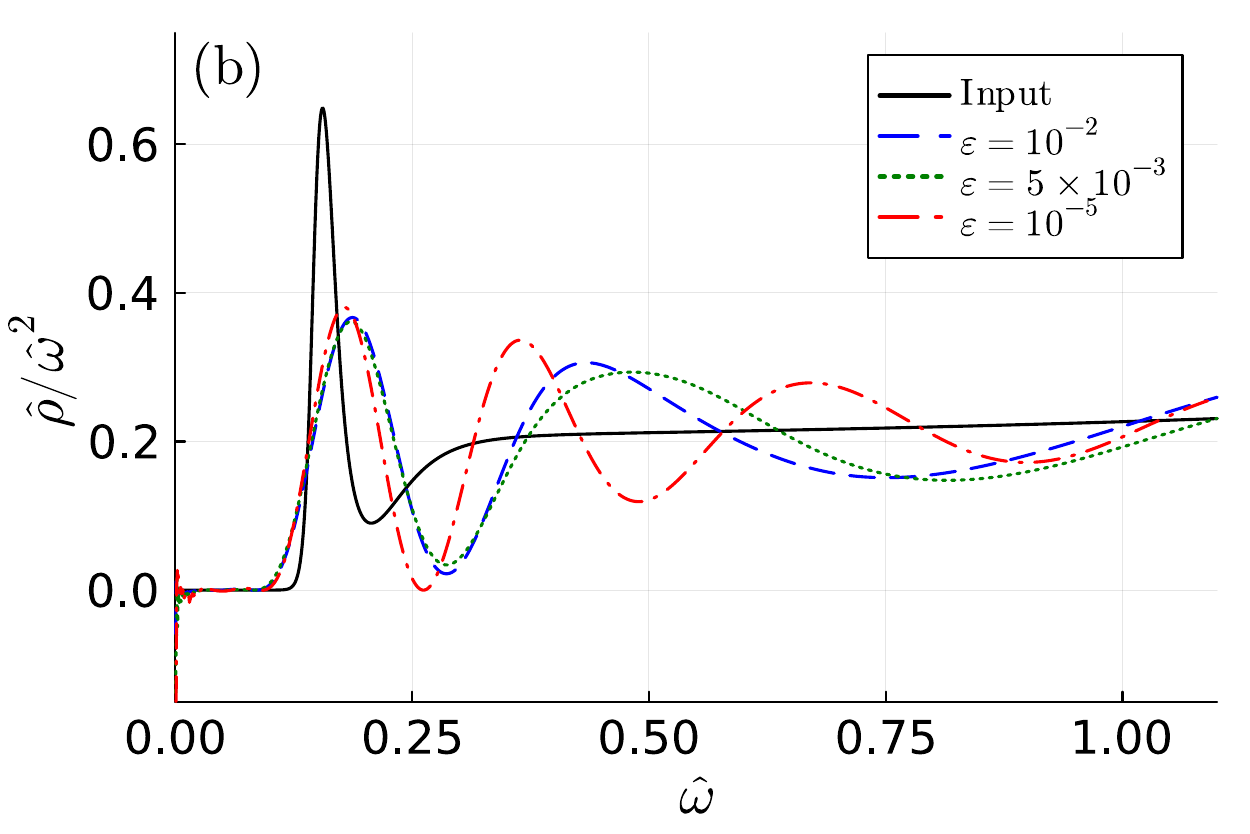}
  \end{minipage}
  \caption{
    Spectral functions obtained by using $K^{(0)}_{W}(\hat{\tau}_{i},\hat{\omega}_{j})$
    in the mock-data tests for $T<T_{\mathrm{c}}$.
    Figure (a) shows the results with a fixed noise level of $\varepsilon=5\times 10^{-3}$.
    The blue dashed, green dotted and red dash-dotted lines 
    represent the output results with $N_{\tau}=48$, 64 and 96,
    respectively.
    Figure (b) shows the results with a fixed temporal extent of $N_{\tau}=96$.
    The blue dashed, green dotted and red dash-dotted lines
    represent the output results with $\varepsilon=10^{-2}$, $5\times 10^{-3}$
    and $10^{-5}$,
    respectively.
    In both figures, 
    the black solid line represents the input spectral function.
  }
  \label{fig:spf0_mock-data_tests_rho_bTc}
\end{figure*}

To check the applicability of SpM
to the number of input data and the magnitude of noise,
we perform tests on three types of temporal extents,
$N_{\tau}=$ 48, 64 and 96,
and three types of noise levels,
$\varepsilon=10^{-5}$, $5\times 10^{-3}$ and $10^{-2}$
in which $\varepsilon=5\times 10^{-3}$ is close to the quality of data of correlation functions
calculated by using lattice QCD.
Two types of kernels,
$K^{(0)}_{W}(\hat{\tau}_{i},\hat{\omega}_{j})$
and $K^{(1)}_{W}(\hat{\tau}_{i},\hat{\omega}_{j})$,
are used at temperature below $T_{\mathrm{c}}$,
and $K^{(0)}_{W}(\hat{\tau}_{i},\hat{\omega}_{j})$ is only used
at temperature above $T_{\mathrm{c}}$ in the test.
The parameters $\mu$ and $\mu^{\prime}$ are introduced in the ADMM algorithm
(see Appendix~\ref{sec:ADMM} in detail).
We set $\mu=\mu^{\prime}=10^{4}$ for the test
in which $K^{(0)}_{W}(\hat{\tau}_{i},\hat{\omega}_{j})$ is used
and set $\mu=\mu^{\prime}=10^{3}$ for the test
in which $K^{(1)}_{W}(\hat{\tau}_{i},\hat{\omega}_{j})$ is used
(see Appendix~\ref{sec:mu_determined} for information on how to determine $\mu$ and $\mu^{\prime}$).
In all of the tests,
the range of $\hat{\omega}$ of spectral functions is set from 0 to 4
and the number of points in $\omega$-space is $N_{\omega}=8001$.
In the test for each combination of $N_{\tau}$ and $\varepsilon$,
$\lambda_{\mathrm{opt}}$ is estimated between $3.5\times 10^{2}$ and $6.6\times 10^{4}$.
\\
\indent
Figure~\ref{fig:spf0_mock-data_tests_rho_bTc} shows
the spectral functions as a function of $\hat{\omega}$
for $T<T_{\mathrm{c}}$.
These are obtained using $K^{(0)}_{W}(\hat{\tau}_{i},\hat{\omega}_{j})$,
that is,
$\hat{\rho}(\hat{\omega}_{j})$ is obtained from $\tilde{\rho}^{(0)}(\hat{\omega}_{j})$
calculated directly in ADMM.
In Fig.~\ref{fig:spf0_mock-data_tests_rho_bTc}(a),
the results with a fixed noise level of $\varepsilon=5\times 10^{-3}$
for various $N_{\tau}$ are illustrated.
The black solid line represents the input spectral function,
and the blue dashed, green dotted and red dash-dotted lines
represent the output results with $N_{\tau}=48$, 64 and 96,
respectively.
It can be seen that
the result for $N_{\tau}=96$ has the sharpest peak
and the position of the peak is closest to the input spectral function.
Similar results are obtained for the other noise levels $\varepsilon$.
This means that
the more data points in the input correlation function,
the better the reproducibility of the spectral function.
In Fig.~\ref{fig:spf0_mock-data_tests_rho_bTc}(b),
the results with a fixed temporal extent of $N_{\tau}=96$
for various $\varepsilon$ are illustrated.
The black solid line represents the input spectral function,
and the blue dashed, green dotted and red dash-dotted lines
represent the output results with $\varepsilon=10^{-2}$, $5\times 10^{-3}$ and $10^{-5}$,
respectively.
It can be seen that
the result for $\varepsilon=10^{-5}$ has the sharpest peak
and the position of the peak is closest to the input spectral function.
Similar results are obtained for the other temporal extent $N_{\tau}$.
This means that
the smaller the errors in the input correlation function,
the higher the reproducibility.
\\
\indent
In Fig.~\ref{fig:spf0_mock-data_tests_rho_bTc},
the spectral function in the vicinity of $\hat{\omega}=0$ exhibits a pronounced change in value,
abruptly deviating to a positive or negative value.
$\tilde{\rho}^{(0)}(\hat{\omega}_{j})$ is directly obtained by using ADMM
and $\hat{\rho}(\hat{\omega}_{j})$ is calculated using Eq.~\eqref{eq:rho_tilde_omega_hat}.
Therefore,
$\hat{\rho}(\hat{\omega}_{j})$ does not have to be a large positive or negative value
for a small value of $\hat{\omega}$,
reflecting the input spectral function
and the positivity condition imposed on the spectral function in ADMM algorithm.
However,
the magnitude of $\hat{\rho}(\hat{\omega}_{j})/\hat{\omega}^{2}$ becomes large
since $\hat{\rho}(\hat{\omega}_{j})$ is divided by the square of a small value of $\hat{\omega}$.
Thus,
what appears to be a peak in the vicinity of $\hat{\omega}=0$ is,
in fact,
a fake.
\begin{figure*}[htb]
  \centering
  \begin{minipage}{.49\textwidth}
    \includegraphics[width=1.0\linewidth]{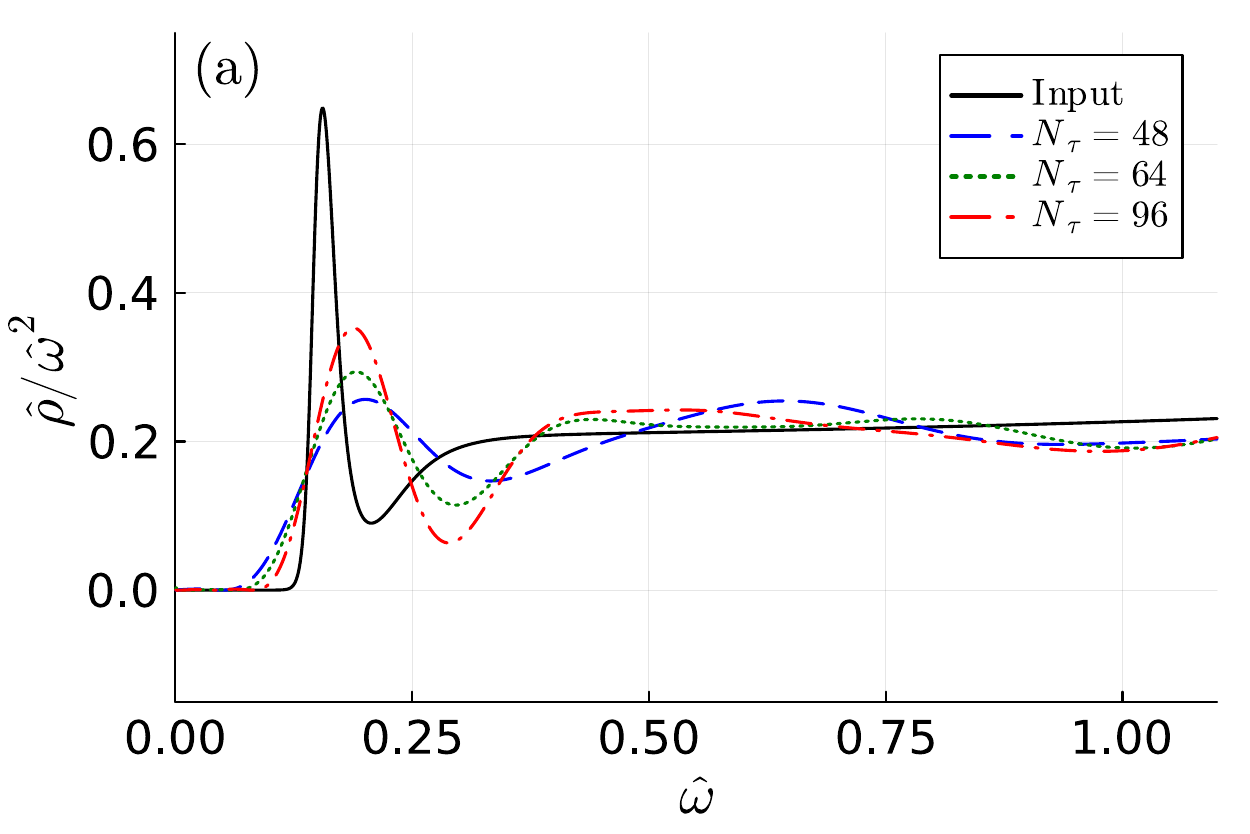}
  \end{minipage}
  \begin{minipage}{.49\textwidth}
    \includegraphics[width=1.0\linewidth]{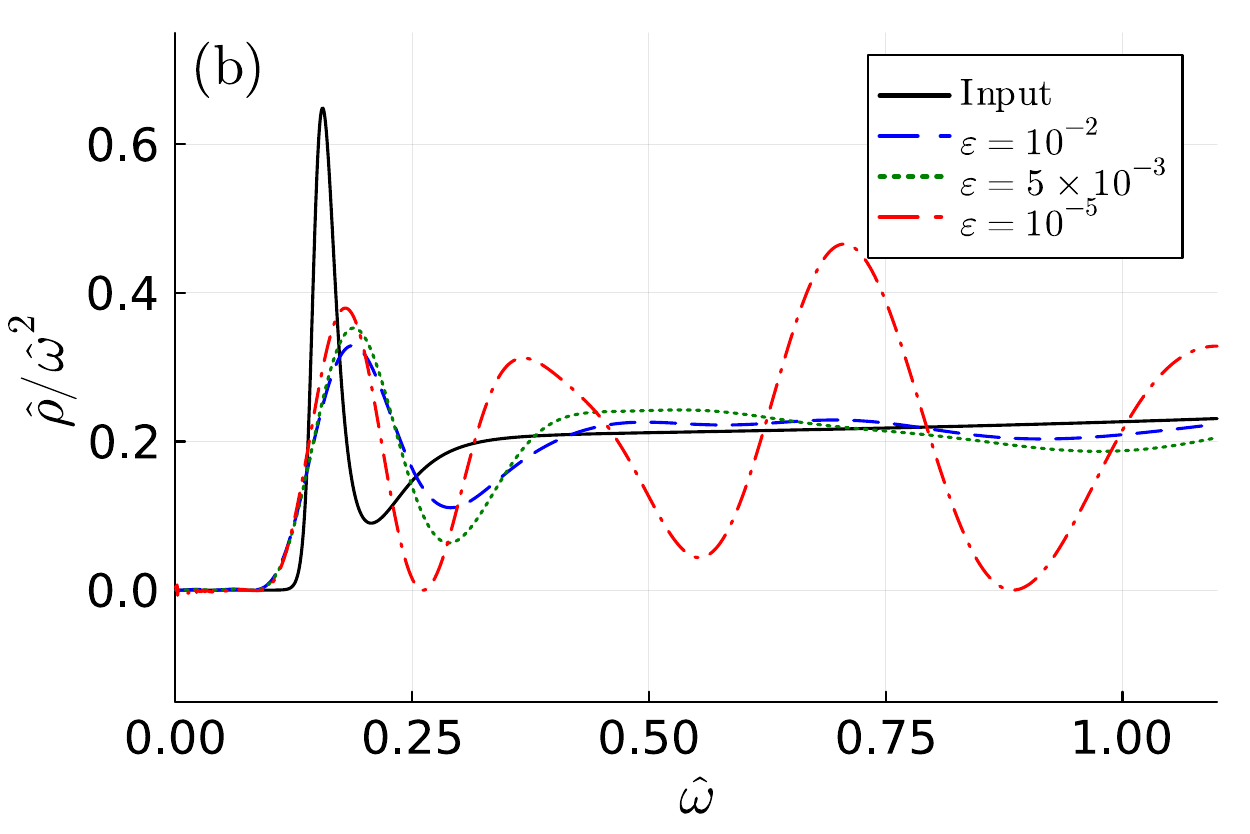}
  \end{minipage}
  \caption{
    Same as Fig.~\ref{fig:spf0_mock-data_tests_rho_bTc}
    but obtained by using $K^{(1)}_{W}(\hat{\tau}_{i},\hat{\omega}_{j})$.
    Figure (a) and (b) show the results
    with $\varepsilon=5\times 10^{-3}$ and $N_{\tau}=96$,
    respectively.
  }
  \label{fig:spf1_mock-data_tests_rho_bTc}
\end{figure*}

\begin{figure*}[tbp]
  \centering
  \begin{minipage}{.49\textwidth}
    \includegraphics[width=1.0\linewidth]{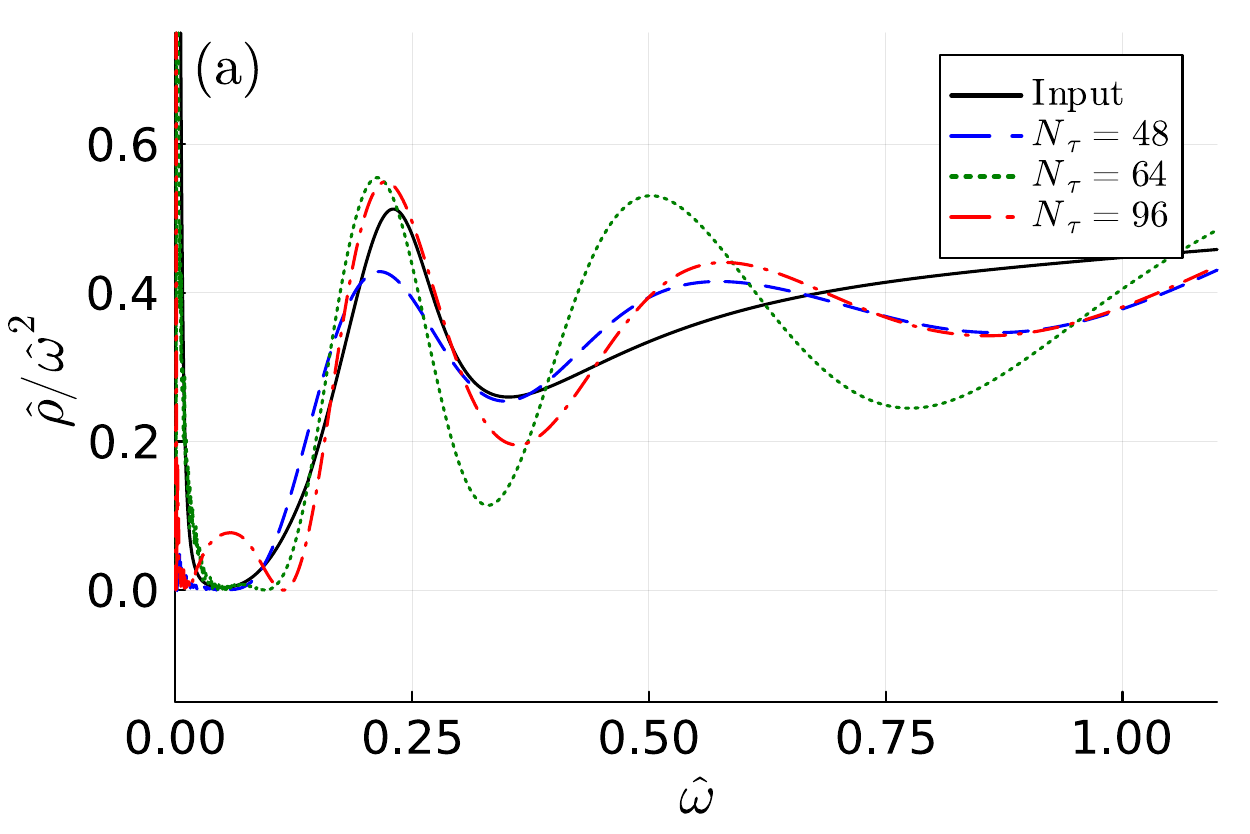}
  \end{minipage}
  \begin{minipage}{.49\textwidth}
    \includegraphics[width=1.0\linewidth]{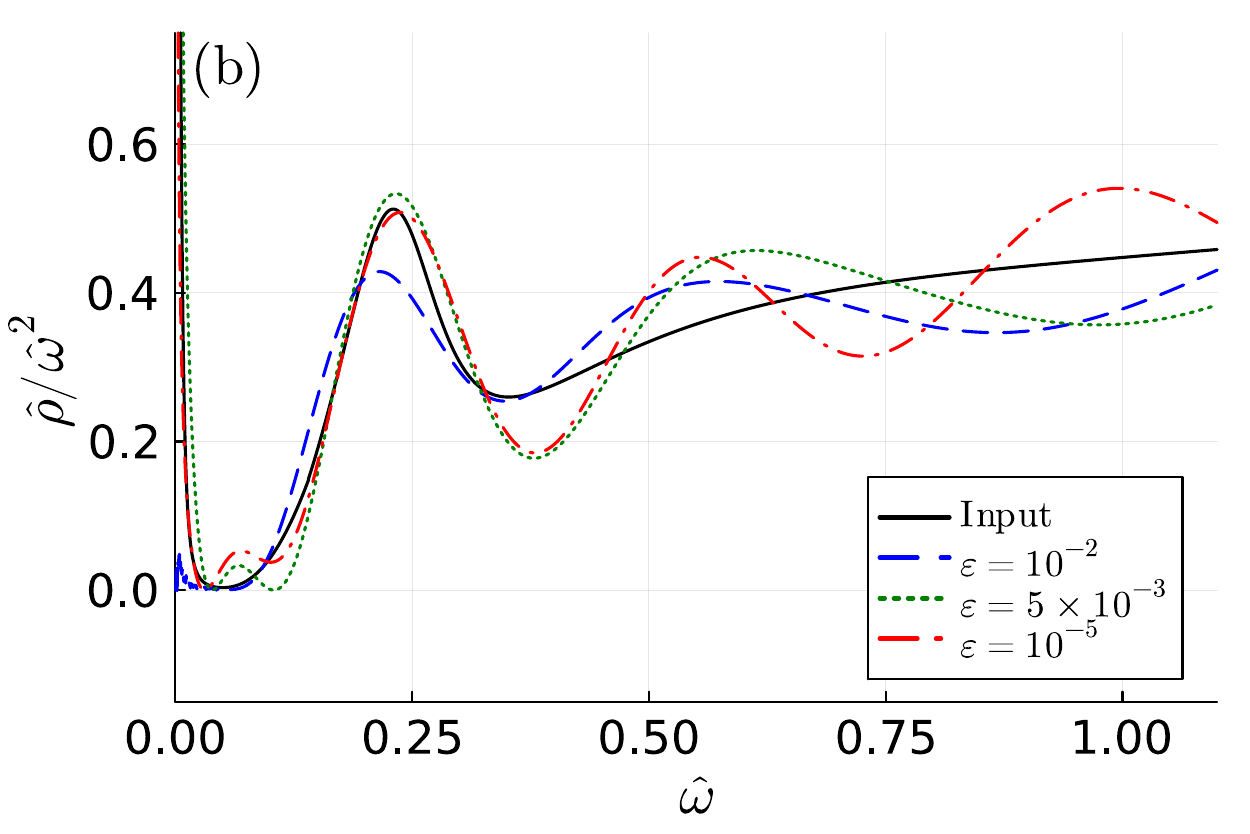}
  \end{minipage}
  \caption{
    Same as Fig.~\ref{fig:spf0_mock-data_tests_rho_bTc}
    but for $T>T_{\mathrm{c}}$.
    Figure (a) and (b) show the results
    with $\varepsilon=10^{-2}$ and $N_{\tau}=48$,
    respectively.
  }
  \label{fig:spf0_mock-data_tests_rho_aTc}
\end{figure*}

\indent
Figure~\ref{fig:spf1_mock-data_tests_rho_bTc} shows
the same results as in Fig.~\ref{fig:spf0_mock-data_tests_rho_bTc}
but obtained by using $K^{(1)}_{W}(\hat{\tau}_{i},\hat{\omega}_{j})$,
that is,
$\hat{\rho}(\hat{\omega}_{j})$ are obtained from $\tilde{\rho}^{(1)}(\hat{\omega}_{j})$
calculated directly in ADMM.
From Fig.~\ref{fig:spf1_mock-data_tests_rho_bTc}(a),
increasing $N_{\tau}$ results in better spectral functions
that are closer to the input spectral function.
Similar results are obtained for the other noise levels $\varepsilon$.
From Fig.~\ref{fig:spf1_mock-data_tests_rho_bTc}(b),
reducing $\varepsilon$ results in better spectral functions
that are closer to the input spectral function.
Similar results are obtained for the other temporal extents $N_{\tau}$.
\\
\indent
In contrast to the results of the spectral functions
with $K^{(0)}_{W}(\hat{\tau}_{i},\hat{\omega}_{j})$
shown in Fig.~\ref{fig:spf0_mock-data_tests_rho_bTc},
the values of the spectral function in the vicinity of $\hat{\omega}=0$ are not large.
As mentioned in Sect.~\ref{subsec:formalism_SpM},
$\tilde{\rho}^{(1)}(0)=0$ is implicitly imposed,
that is,
$\hat{\rho}(0)=0$ is ensured.
Thus,
there would be no fake peak.
\\
\indent
Figure~\ref{fig:spf0_mock-data_tests_rho_aTc} shows
the same results as in Fig.~\ref{fig:spf0_mock-data_tests_rho_bTc},
but for $T>T_{\mathrm{c}}$.
From Fig.~\ref{fig:spf0_mock-data_tests_rho_aTc}(a),
it can be seen that
the result for $N_{\tau}=96$ has the closest resonance peak to the input spectral function.
Similar results are obtained for the other noise levels $\varepsilon$.
This means that
the more data points in the input correlation function,
the better the reproducibility of the spectral function.
From Fig.~\ref{fig:spf0_mock-data_tests_rho_aTc}(b),
it can be seen that
the result for $\varepsilon=10^{-5}$ has the closest resonance peak to the input spectral function.
Similar results are obtained for the other temporal extent $N_{\tau}$.
This means that
the smaller the errors in the input correlation function,
the higher the reproducibility.

\begin{figure}[htbp]
  \centering
  \includegraphics[width=1.0\linewidth]{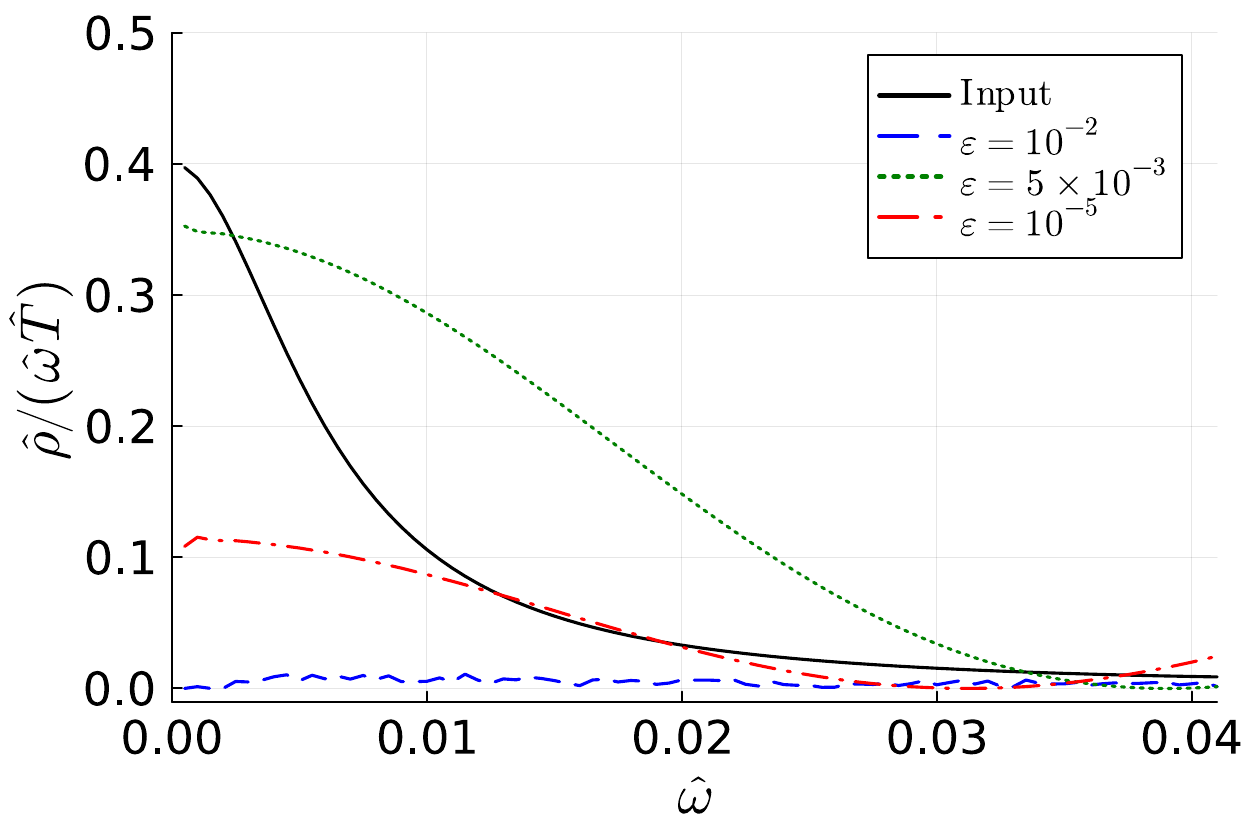}
  \caption{
    $\hat{\rho}/(\hat{\omega}\hat{T})$ as a function of $\hat{\omega}$,
    which is an enlarged view of the plot in Fig.~\ref{fig:spf0_mock-data_tests_rho_aTc}(b)
    in the very small value of $\hat{\omega}$.
    Note that $\hat{T}=N_{\tau}^{-1}$.
  }
  \label{fig:spf0_mock-data_tests_rho_aTc_tpp}
\end{figure}

\indent
The change in value of the spectral function shown in Fig.~\ref{fig:spf0_mock-data_tests_rho_aTc}
is more gradual than for $T<T_{\mathrm{c}}$ shown in Fig.~\ref{fig:spf0_mock-data_tests_rho_bTc}.
In the case of $N_{\tau}=64$ shown in Fig.~\ref{fig:spf0_mock-data_tests_rho_aTc}(a)
and that of $\varepsilon=10^{-5}$ shown in Fig.~\ref{fig:spf0_mock-data_tests_rho_aTc}(b),
these appear to fit well with the transport peak of the input spectral function.
In order to confirm this,
we illustrate $\hat{\rho}/(\hat{\omega}\hat{T})$ as a function of $\hat{\omega}$
in Fig.~\ref{fig:spf0_mock-data_tests_rho_aTc_tpp},
which is an enlarged view of the plot in Fig.~\ref{fig:spf0_mock-data_tests_rho_aTc}(b)
at small $\hat{\omega}$.
Note that $\hat{T}=N_{\tau}^{-1}$.
The values of the spectral function in this region are non-zero,
which may reflect the transport peak in the input spectral function.
On the other hand,
the value of the spectral function in the limit of $\hat{\omega}\to 0$
for $\varepsilon=5\times 10^{-3}$ is closest to that of the input spectral function.
This does not mean that the smaller the error,
the better the reproducibility,
as is the case with the resonance peak.
Even when other combinations of $N_{\tau}$ and $\varepsilon$ are examined,
it cannot be concluded that
a smaller error necessarily leads to better reproducibility of the value of the spectral function
in the limit of $\hat{\omega}\to 0$,
nor that increasing $N_{\tau}$ improves the reproducibility in the same limit.

\indent
The results for $T>T_{\mathrm{c}}$ shown in Fig.~\ref{fig:spf0_mock-data_tests_rho_aTc} indicate that
some of the input and output resonance peaks are relatively well matched,
whereas the results for $T<T_{\mathrm{c}}$ shown
in Figs.~\ref{fig:spf0_mock-data_tests_rho_bTc} and~\ref{fig:spf1_mock-data_tests_rho_bTc} indicate that
they are not well matched to the sharp peaks in the input spectral function.
The transport peak mentioned above is also accompanied by an abrupt change in value.
These results suggest that
sparse solutions of spectral functions in the IR basis
may not adequately represent the abrupt change in value of the spectral function in the original basis.
\\
\indent
In regions where $\hat{\omega}\gtrapprox 0.30$,
the output spectral function has wavy results.
In particular,
spectral functions with larger amplitudes tend to be obtained
when the noise level $\varepsilon$ is small.
It can be seen that these results show the overfitting behavior for the spectral function in the regions.
It is shown in~\cite{Otsuki.PhysRevE.95.061302} that
SpM with weak regularization
($\lambda<\lambda_{\mathrm{opt}}$)
results in overfitting.
Although $\lambda_{\mathrm{opt}}$ is decided following the previous study,
it appears that when the error in the correlation function is small
and the spectral function has a sharp peak structure,
$\lambda_{\mathrm{opt}}$ is chosen,
which is overfitting for the spectral function in the regions.
On the basis of the above discussion,
an additional assumption beyond sparseness may be needed
for extracting a spectral function with an abrupt change in value.
\\
\indent
We also conduct several tests using mock data of spectral functions with two resonance peaks in the low-frequency region.
The results confirm that only the output of the first peak is valid,
and that the positions of the second and subsequent peaks cannot be determined.
Furthermore, when the mock spectral function has two sharp peaks within a narrow frequency range,
the peak obtained by SpM in that frequency range appears as a single broad peak.
As mentioned in Sec~\ref{subsec:formalism_SpM} section,
SpM preferentially reconstructs spectral functions with moderate width, as sharply localized peaks typically involve contributions from high-frequency modes that are poorly constrained by the Euclidean correlator data.
As a result, reconstructions based solely on the sparsity assumption tend to yield spectral functions with a resolution limited by the information content of the data.

\subsection{
Validation using free spectral functions
\label{subsec:validation_spf}
}
\begin{figure*}[tb]
  \centering
  \begin{minipage}{.49\textwidth}
    \includegraphics[width=1.0\linewidth]{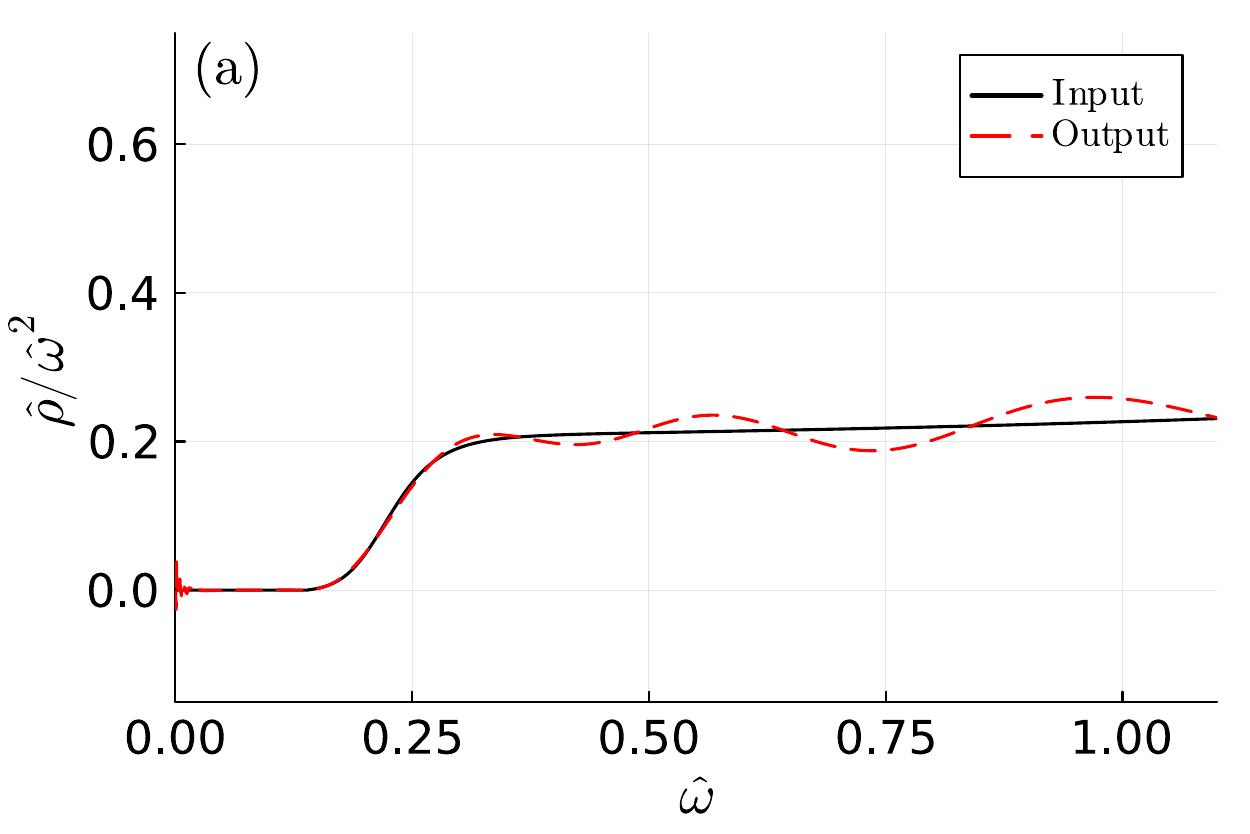}
  \end{minipage}
  \begin{minipage}{.49\textwidth}
    \includegraphics[width=1.0\linewidth]{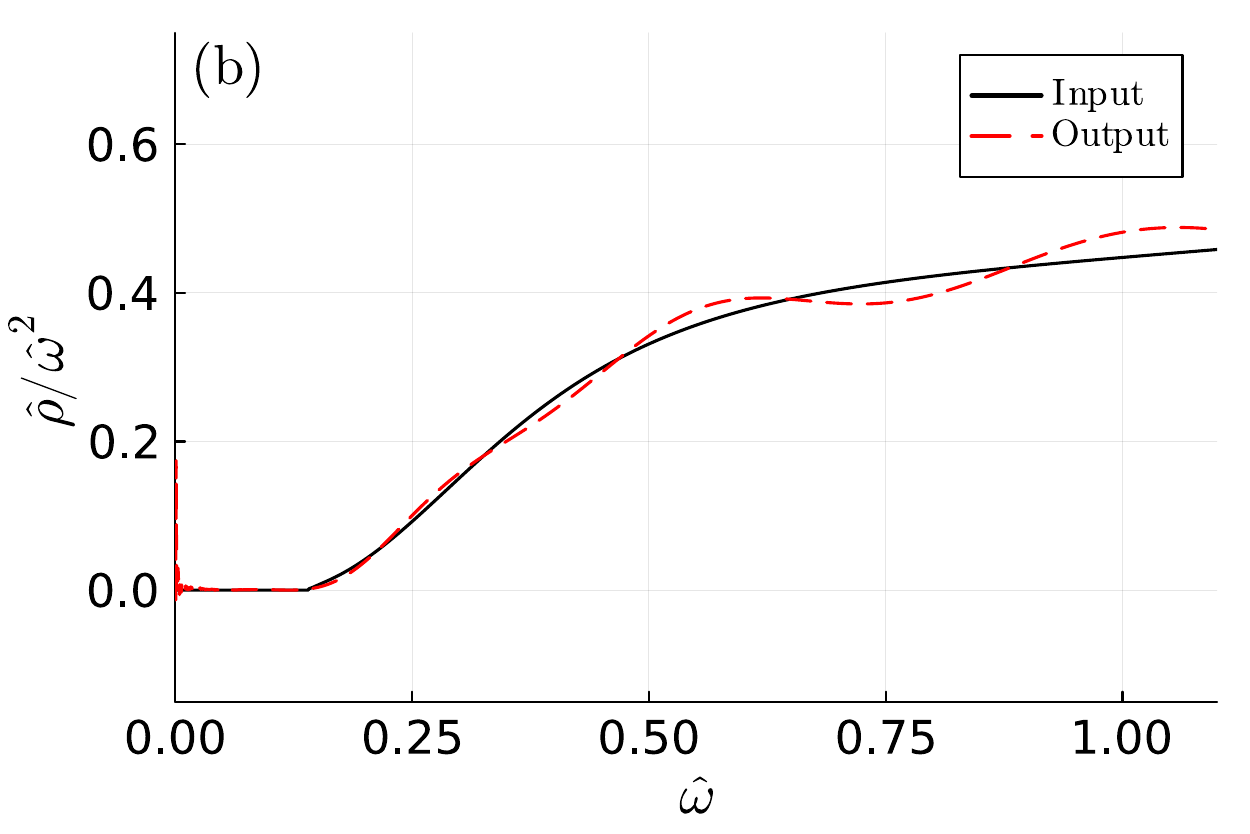}
  \end{minipage}
  \caption{
    Spectral functions reconstructed from correlation functions calculated using the mock free spectral function.
    Figure (a) shows the results with $K^{(0)}_{W}(\hat{\tau}_{i},\hat{\omega}_{j})$, $N_{\tau}=96$ and a noise level of $\varepsilon=10^{-5}$ for $T<T_{\mathrm{c}}$.
    Figure (b) shows the results with $K^{(0)}_{W}(\hat{\tau}_{i},\hat{\omega}_{j})$, $N_{\tau}=48$ and a noise level of $\varepsilon=10^{-5}$ for $T>T_{\mathrm{c}}$.
    In both figures, 
    the black solid and the red dashed lines represent the input and the output spectral function, respectively.
  }
  \label{fig:spf0_free_spf_mock-data_tests}
\end{figure*}

In order to further assess the reliability of the spectral functions reconstructed by SpM,
as discussed in the previous section,
we conduct additional tests using mock spectral functions that contain no peaks.
We have extracted spectral functions by using SpM from correlation functions
made from mock spectral functions with no peaks,
and the results of the output spectral function match the input spectral function well~\cite{Takahashi:2023Fc}.
However,
since the settings considered in~\cite{Takahashi:2023Fc} differ from those in this study,
it is meaningful to revisit investigations using spectral functions without peaks.
\\
\indent
The input spectral functions without peaks at temperatures below and above $T_{\mathrm{c}}$ are prepared based on the mock spectral functions introduced in Sec.~\ref{subsec:mock_spf}.
For temperatures below $T_{\mathrm{c}}$,
we use the spectral function with $c_{\mathrm{res}}=0$,
consisting solely of the free Wilson spectral function and the modified $\Theta$ function.
For temperatures above $T_{\mathrm{c}}$,
we adopt the spectral function with $c_{\mathrm{res}}=c_{\mathrm{trans}}=0$,
again consisting only of the free Wilson spectral function and the modified $\Theta$ function.
Note that all other settings— such as $N_{\tau}$, $\varepsilon$, $\mu$, and $\mu^{\prime}$ —are kept identical to those used in the previous section.
\\
\indent
Figure~\ref{fig:spf0_free_spf_mock-data_tests} shows our typical results of the spectral functions.
In Fig.~\ref{fig:spf0_free_spf_mock-data_tests}(a),
the red dashed line represents the output result with $N_{\tau}=96$ and $\varepsilon=10^{-5}$ for $T<T_{\mathrm{c}}$,
and the black solid line represents the input spectral function.
In Fig.~\ref{fig:spf0_free_spf_mock-data_tests}(b),
the red dashed line represents the output result with $N_{\tau}=48$ and $\varepsilon=10^{-5}$ for $T>T_{\mathrm{c}}$,
and the black solid line represents the input spectral function.
The output results at both temperatures show good agreement with their respective inputs.
Similar results are obtained across various combinations of $N_{\tau}$ and $\varepsilon$.
Furthermore,
note that the results were obtained by using $K^{(0)}_{W}(\hat{\tau}_{i},\hat{\omega}_{j})$,
and we confirm that results of comparable quality are also obtained when using $K^{(1)}_{W}(\hat{\tau}_{i},\hat{\omega}_{j})$.
\\
\indent
The input free spectral function exhibits different frequency dependences below and above $T_{\mathrm{c}}$,
yet SpM reconstructs these behaviors successfully.
The results presented in Sec.~\ref{subsec:dependence_Nt_epsilon} demonstrate that the output of SpM reflects the increase in values due to the presence of resonance peaks in the spectral function.
Conversely, when the input correlation function lacks information of a resonance peak,
SpM does not generate spurious peak structures.

\section{
Results from lattice QCD data
\label{sec:results_LQCD}
}
\begin{figure*}[htb]
  \centering
  \begin{minipage}{.49\textwidth}
    \includegraphics[width=1.0\linewidth]{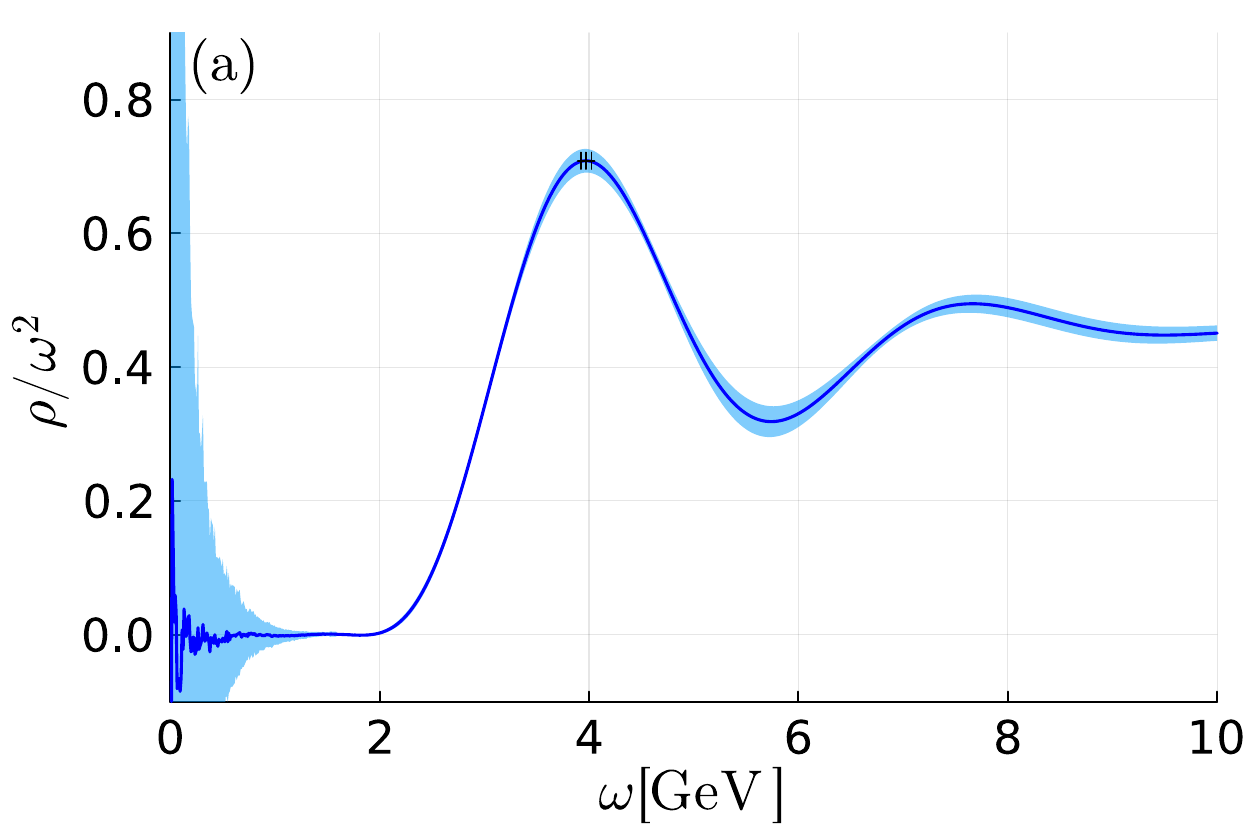}
  \end{minipage}
  \begin{minipage}{.49\textwidth}
    \includegraphics[width=1.0\linewidth]{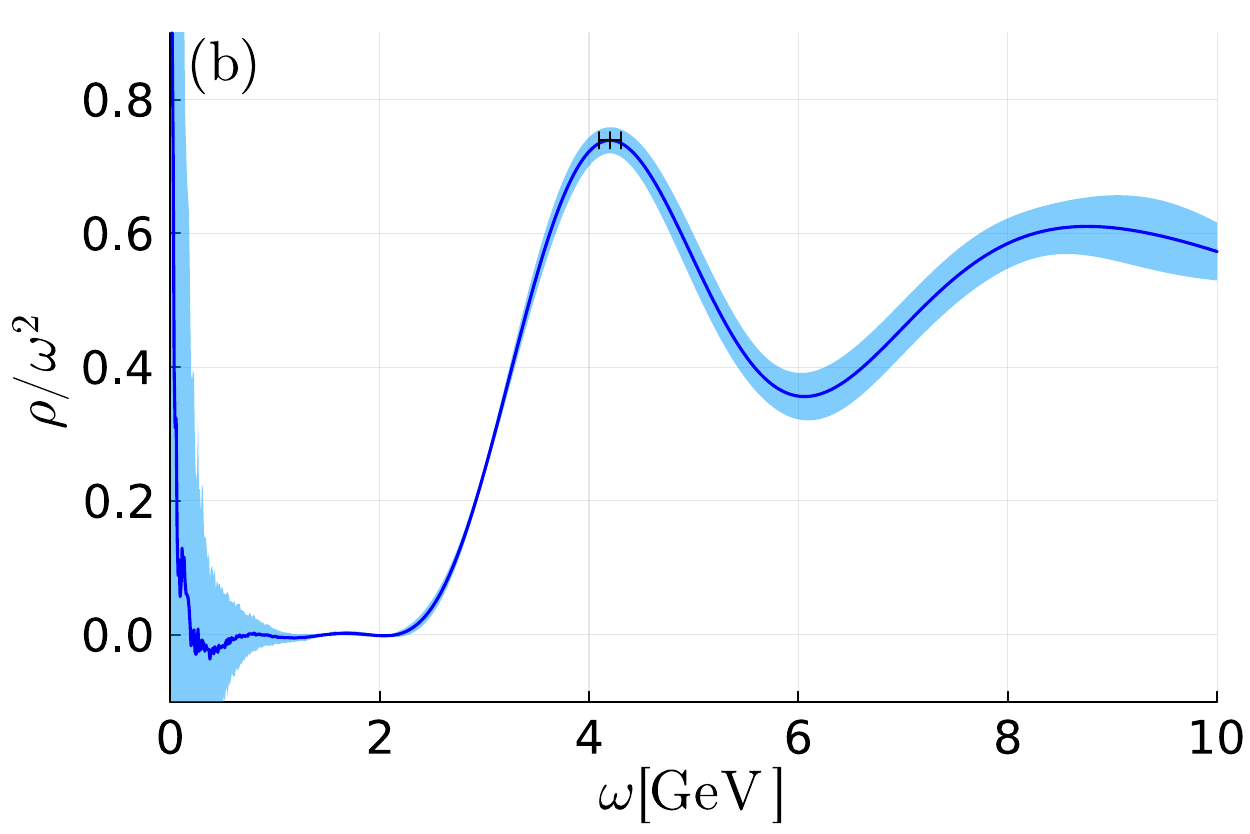}
  \end{minipage}
  \caption{
    Spectral functions obtained by using $K^{(0)}_{W}(\hat{\tau}_{i},\hat{\omega}_{j})$
    in (a) pseudoscalar and (b) vector channels
    extracted from actual lattice QCD data for $T\simeq 0.73T_{\mathrm{c}}$.
    The parameters $\mu$ and $\mu^{\prime}$ in ADMM algorithm
    are set to $10^{2}$.
    The blue shaded areas represent the statistical errors of the spectral functions
    from Jackknife analyses,
    the blue solid lines represent the mean values,
    and the black horizontal error bars represent
    the uncertainty of the location of the first peak for each spectral function.
  }
  \label{fig:spf_LQCD_0.73Tc}
\end{figure*}

In this section
we show results of the charmonium spectral function
in pseudoscalar and vector channels extracted
from the correlation functions calculated by using lattice QCD.
These lattice data used in this study were given in ref.~\cite{Ding2012.PhysRevD.86.014509},
where the correlation functions were measured with the $O(a)$-improved Wilson quark action
on quenched gauge configurations generated by using the standard plaquette gauge action.
The lattice spacing $a=0.010$ fm
and the corresponding $a^{-1}$ is about 18.97 GeV.
The spatial extent $N_{\sigma}$ is 128,
and the temporal extent $N_{\tau}$ is 48 and 96.
These setup correspond to temperature $T\simeq 0.73T_{\mathrm{c}}$
and $1.46T_{\mathrm{c}}$.
The number of gauge configurations is 234 for $T\simeq 0.73T_{\mathrm{c}}$
and 461 for $T\simeq 1.46T_{\mathrm{c}}$.
We utilized meson correlation functions
in the pseudoscalar and the vector channels
for each temperature.
We set $\hat{\tau}_{\mathrm{r}}=4$,
and use the data of the correlation function from $\hat{\tau}_{\mathrm{r}}$ to $N_{\tau}/2$.
\\
\indent
As in the case of mock-data tests,
two types of kernel written in Eq.~\eqref{eq:kernel_W} are considered in our calculations,
that is,
$K^{(0)}_{W}(\hat{\tau}_{i},\hat{\omega}_{j})$
and $K^{(1)}_{W}(\hat{\tau}_{i},\hat{\omega}_{j})$
are used for $T\simeq 0.73T_{\mathrm{c}}$,
and only $K^{(0)}_{W}(\hat{\tau}_{i},\hat{\omega}_{j})$ is used for $T\simeq 1.46T_{\mathrm{c}}$.
We also set the range of $\hat{\omega}$ of the spectral functions from 0 to 4,
and the number of points in $\omega$-space to $N_{\omega}=8001$.
In the analyses using actual lattice QCD data,
$\lambda_{\mathrm{opt}}$ is estimated between $4.0\times 10^{3}$ and $3.1\times 10^{5}$.
The values of the parameters $\mu$ and $\mu^{\prime}$ are set
$\mu=\mu^{\prime}=10^{2}$, $10^{3}$ and $10^{4}$
(see Appendix~\ref{sec:mu_determined} for information on how to determine $\mu$ and $\mu^{\prime}$).
The statistical error of the spectral function is estimated using the Jackknife method with a bin size of one.
\\
\indent
Figure~\ref{fig:spf_LQCD_0.73Tc} shows
our typical results of the spectral functions
in (a) vector and (b) pseudoscalar channels for $T\simeq 0.73T_{\mathrm{c}}$.
These are obtained using $K^{(0)}_{W}(\hat{\tau}_{i},\hat{\omega}_{j})$,
that is,
$\hat{\rho}(\hat{\omega}_{j})$ is obtained from $\tilde{\rho}^{(0)}(\hat{\omega}_{j})$
calculated directly in ADMM,
and the parameters $\mu$ and $\mu^{\prime}$ in ADMM algorithm are set to $10^{2}$.
The blue shaded areas represent the statistical errors of the spectral functions
from Jackknife analyses,
the blue solid lines represent the mean values,
and the black horizontal error bars at the top of the first peaks represent
the uncertainty of the location of the resonance peak for each spectral function.
The value of the spectral function increases around 2 GeV,
and the average value of the location of the first peak is about 4 GeV in both channels.
Similar results are obtained for $\mu=\mu^{\prime}=10^{3}$ and $10^{4}$.
The mean values,
statistical errors,
and systematic errors of the location of the resonance peak,
which are estimated after considering all values of $\mu$,
are listed in Table~\ref{tbl:spf_mean-err_LQCD}.
The mean values are 4.02 GeV in the pseudoscalar channel
and 4.30 GeV in the vector channel,
while those obtained from MEM
are 3.31 GeV in the pseudoscalar channel
and 3.48 GeV in the vector channel~\cite{Ding2012.PhysRevD.86.014509}.
Although our result is somewhat larger than that of the previous study,
the qualitative behavior is consistent: a well-defined resonance peak is observed,
with its position in the pseudoscalar channel appearing at a lower frequency
than in the vector channel.

\begin{figure*}[tbp]
  \centering
  \begin{minipage}{.49\textwidth}
    \includegraphics[width=1.0\linewidth]{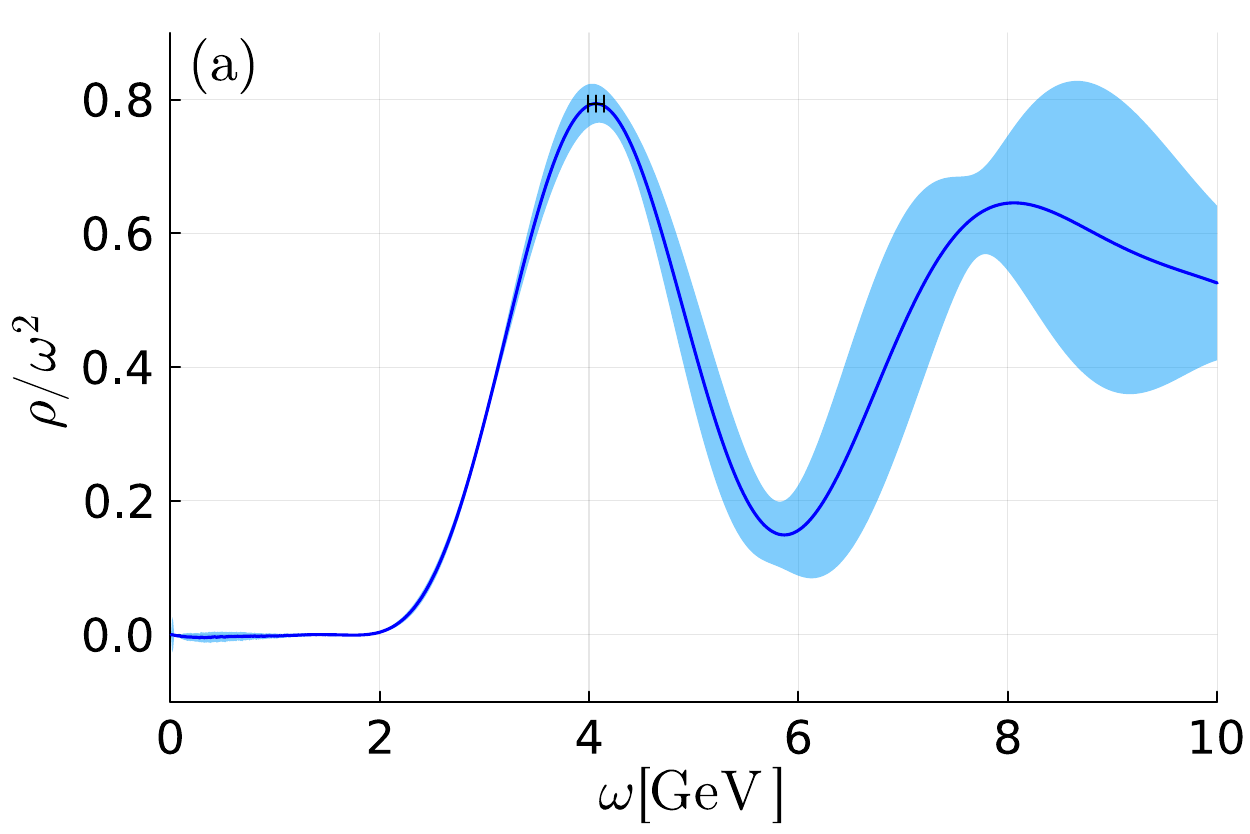}
  \end{minipage}
  \begin{minipage}{.49\textwidth}
    \includegraphics[width=1.0\linewidth]{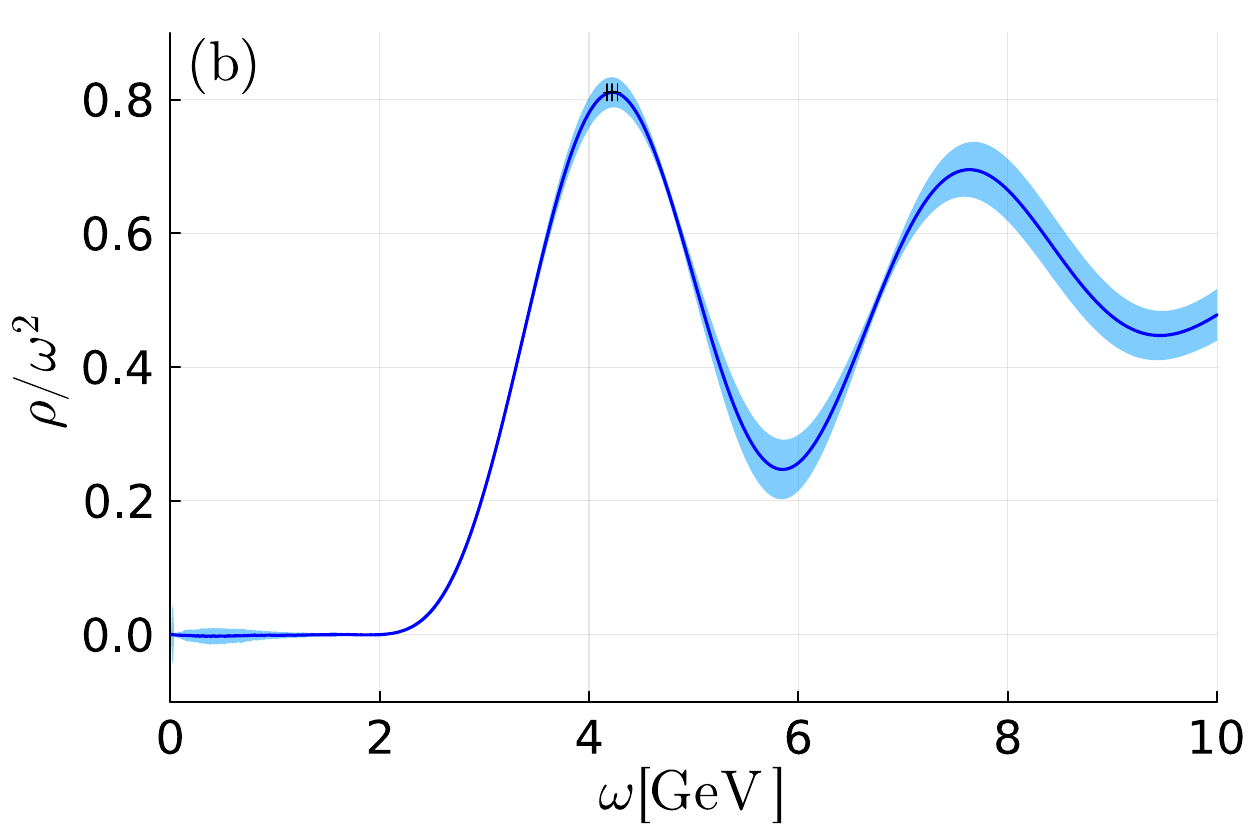}
  \end{minipage}
  \caption{
    Same as Fig.~\ref{fig:spf_LQCD_0.73Tc}
    but obtained by using $K^{(1)}_{W}(\hat{\tau}_{i},\hat{\omega}_{j})$.
    Figure (a) and (b) show the results
    with pseudoscalar and vector channels,
    respectively.
    The parameters $\mu$ and $\mu^{\prime}$ in ADMM algorithm are set to $10^{2}$.
  }
  \label{fig:spf_LQCD_0.73Tc_rho-o-w}
\end{figure*}

\indent
Focusing on the very low frequency of the spectral function shown in Fig.~\ref{fig:spf_LQCD_0.73Tc},
we can see that it slightly fluctuates around $\rho=0$ in the pseudoscalar channel
or rapidly increases in value in the vector channel.
It is similar to the spectral functions
obtained using $K^{(0)}_{W}(\hat{\tau}_{i},\hat{\omega}_{j})$ in the mock-data tests.
Since $\hat{\rho}(\hat{\omega}_{j})$ is calculated using Eq.~\eqref{eq:rho_tilde_omega_hat},
$\rho(\omega)$ does not have to be a large magnitude for a small value of $\omega$.
Since $\rho(\omega)$ is divided by the square of a small value of $\omega$,
however,
the magnitude of $\rho(\omega)/\omega^{2}$ becomes large.
The statistical errors represented by the blue shaded areas are also very large.
This means that
the spectral function at low frequencies
calculated using $K^{(0)}_{W}(\hat{\tau}_{i},\hat{\omega}_{j})$
has large differences for each Jackknife sample
and its value is not settled.
At a temperature below $T_{\mathrm{c}}$,
$\rho=0$ is expected in the limit of $\omega\to 0$,
i.e.,
it may be better not to use $K^{(0)}_{W}(\hat{\tau}_{i},\hat{\omega}_{j})$
in calculations at a temperature below $T_{\mathrm{c}}$.
\\
\indent
Figure~\ref{fig:spf_LQCD_0.73Tc_rho-o-w} shows
the same results as in Fig.~\ref{fig:spf_LQCD_0.73Tc}
but obtained using $K^{(1)}_{W}(\hat{\tau}_{i},\hat{\omega}_{j})$.
The results are similar to those
obtained using $K^{(0)}_{W}(\hat{\tau}_{i},\hat{\omega}_{j})$,
and similar results are obtained for $\mu=\mu^{\prime}=10^{3}$ and $10^{4}$.
As shown in the values of the resonance peak listed in Table~\ref{tbl:spf_mean-err_LQCD},
the position of the resonance peak
from $\tilde{\rho}^{(0)}(\hat{\omega}_{j})$
is almost the same as that
from $\tilde{\rho}^{(1)}(\hat{\omega}_{j})$.
Therefore,
the qualitative behavior remains consistent
with the previous study~\cite{Ding2012.PhysRevD.86.014509}.
This common behavior strongly suggests that it reflects the underlying physics of the charmonium spectral function.

In contrast to the results of the spectral functions
with $K^{(0)}_{W}(\hat{\tau}_{i},\hat{\omega}_{j})$
shown in Fig.~\ref{fig:spf_LQCD_0.73Tc},
the average values of the spectral function in the low-$\omega$ region are flat at $\rho=0$.
It is similar to the spectral functions
obtained using $K^{(1)}_{W}(\hat{\tau}_{i},\hat{\omega}_{j})$ in the mock-data tests.
Since $\hat{\rho}(0)=0$ is implicitly imposed,
$\rho(0)=0$ is ensured.
In addition,
the statistical errors of the spectral function in the low-$\omega$ region are very small.
This means that
the spectral function in low-$\omega$ region
calculated using $K^{(1)}_{W}(\hat{\tau}_{i},\hat{\omega}_{j})$
differs little for each Jackknife sample.
Therefore,
SpM analysis using $K^{(1)}_{W}(\hat{\tau}_{i},\hat{\omega}_{j})$ gives a stable solution.

\begin{figure*}[tbp]
  \centering
  \begin{minipage}{.49\textwidth}
    \includegraphics[width=1.0\linewidth]{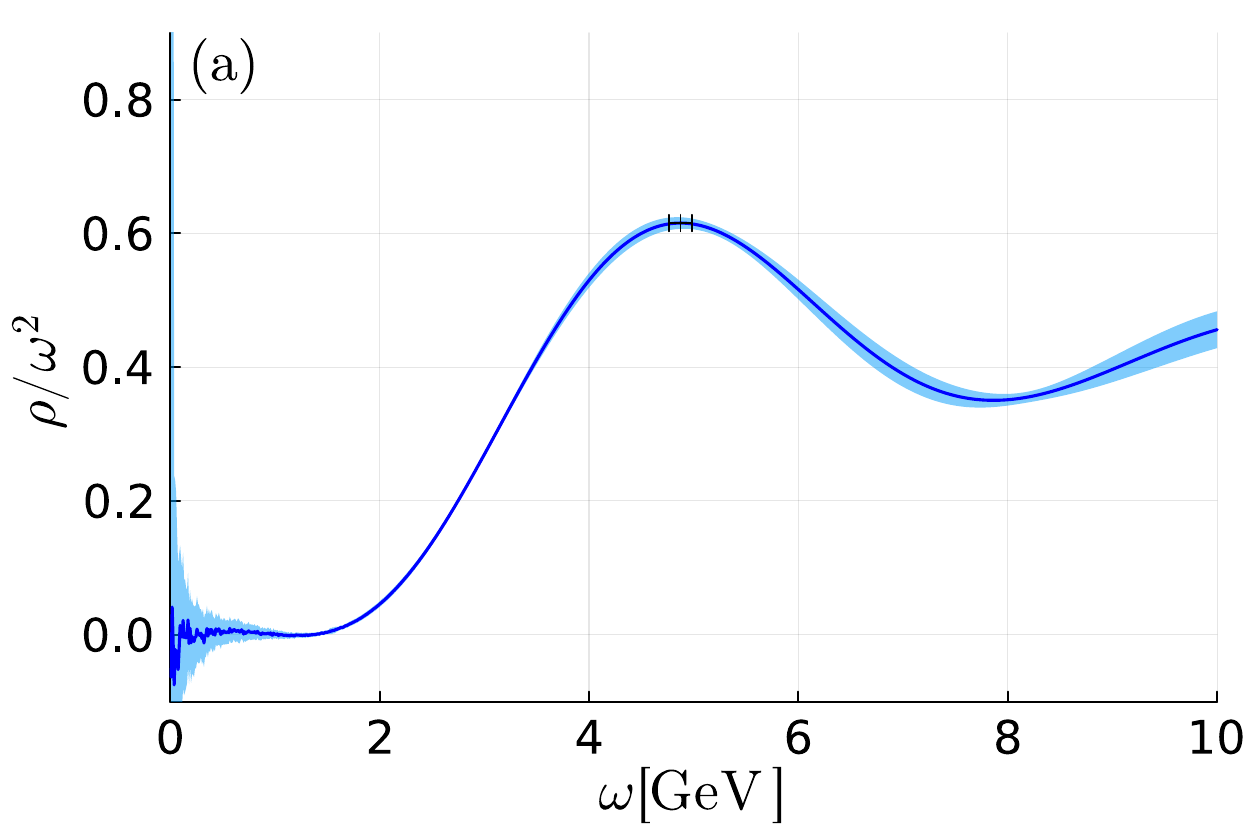}
  \end{minipage}
  \begin{minipage}{.49\textwidth}
    \includegraphics[width=1.0\linewidth]{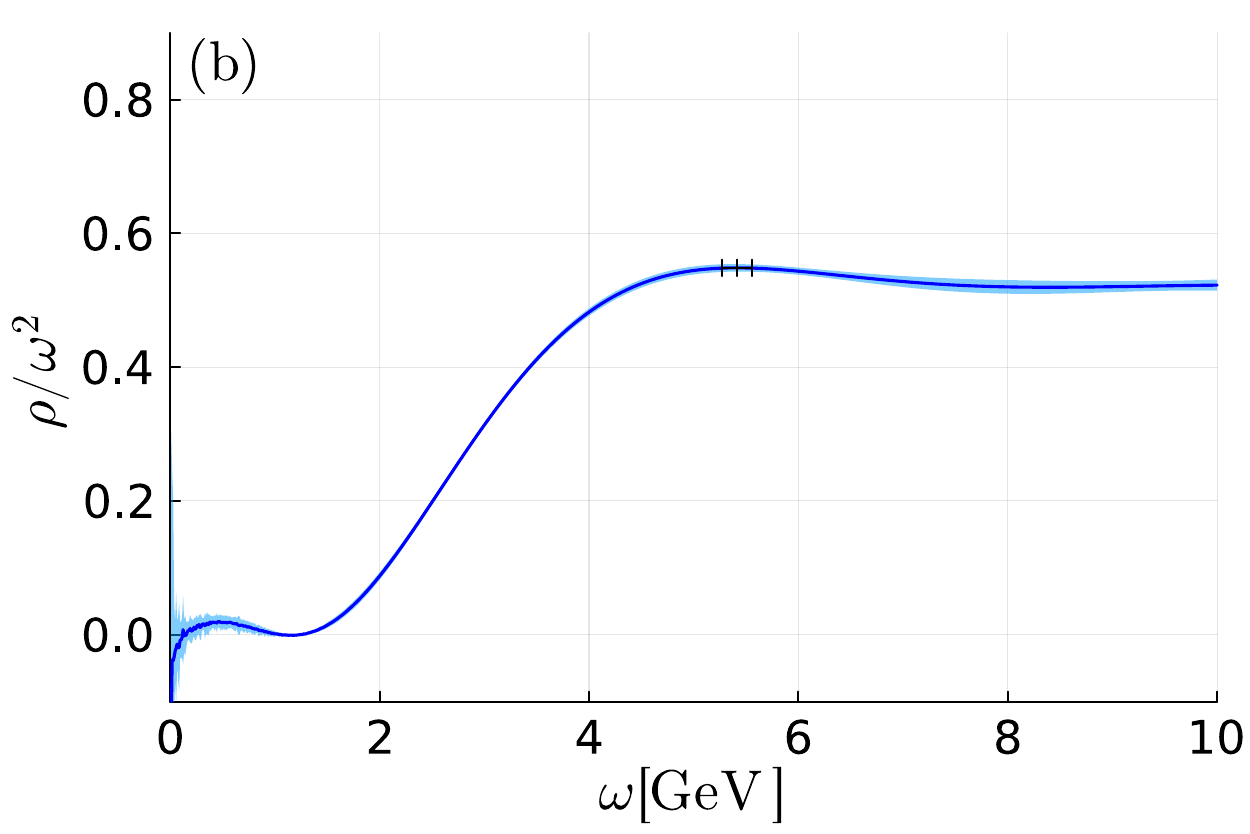}
  \end{minipage}
  \caption{
    Same as Fig.~\ref{fig:spf_LQCD_0.73Tc}
    but for $T\simeq 1.46T_{\mathrm{c}}$.
    Figure (a) and (b) show the results in pseudoscalar and vector channels,
    respectively.
    The parameters $\mu$ and $\mu^{\prime}$ in ADMM algorithm
    are set to $10^{2}$.
  }
  \label{fig:spf_LQCD_1.46Tc}
\end{figure*}

\indent
Figure~\ref{fig:spf_LQCD_1.46Tc} shows
the same results as in Fig.~\ref{fig:spf_LQCD_0.73Tc} but for $T\simeq 1.46T_{\mathrm{c}}$.
Compared to the results for $T\simeq 0.73T_{\mathrm{c}}$,
the peaks are broader and located at higher energies,
and similar results are obtained for $\mu=\mu^{\prime}=10^{2}$ and $10^{4}$.
As shown in the values of the resonance peak listed in Table~\ref{tbl:spf_mean-err_LQCD},
the average values of the location of the resonance peak are 4.87 GeV in the pseudoscalar channel
and 5.42 GeV in the vector channel.
These values are also larger than the results of the previous study~\cite{Ding2012.PhysRevD.86.014509}
in which the results obtained from MEM are about 4.1 GeV in the pseudoscalar channel
and about 4.7 GeV in the vector channel.
However,
the qualitative behavior remains consistent, having a broad peak
with its location in the pseudoscalar channel
appearing at a lower frequency than in the vector channel.
Although systematic errors were not estimated using the same method as in the previous study,
it seems that,
based on the peak width and qualitative consistency with the previous study,
it is possible to conclude that the $\eta_{c}$ corresponding to the pseudoscalar channel and the $J/\psi$ corresponding to the vector channel are already melted at 1.46 $T_{\mathrm{c}}$.

\renewcommand{\arraystretch}{1.3}
\begin{table*}[htb]
\centering
\begin{tabular}{|c|c|c|c|}
\hline
$T/T_c$ & Channel & Kernel $K^{(n)}_{W}$ & Resonance peak \\
\hline
\multirow{4}{*}{0.73} & \multirow{2}{*}{PS} & $n=0$ & $4.02 \pm 0.08\,(\mathrm{stat.})^{+0.04}_{-0.05}\,(\mathrm{sys.})$ \\
\cline{3-4}
 &  & $n=1$ & $4.07 \pm 0.06\,(\mathrm{stat.})^{+0.06}_{-0.06}\,(\mathrm{sys.})$ \\
\cline{2-4}
 & \multirow{2}{*}{VC} & $n=0$ & $4.30 \pm 0.09\,(\mathrm{stat.})^{+0.07}_{-0.10}\,(\mathrm{sys.})$ \\
\cline{3-4}
 &  & $n=1$ & $4.27 \pm 0.07\,(\mathrm{stat.})^{+0.08}_{-0.05}\,(\mathrm{sys.})$ \\
\hline
\multirow{2}{*}{1.46} & PS & $n=0$ & $4.87 \pm 0.08\,(\mathrm{stat.})^{+0.01}_{-0.01}\,(\mathrm{sys.})$ \\
\cline{2-4}
 & VC & $n=0$ & $5.42 \pm 0.16\,(\mathrm{stat.})^{+0.00}_{-0.01}\,(\mathrm{sys.})$ \\
\hline
\end{tabular}
\caption{
The mean values,
statistical errors,
and systematic errors of the location of the resonance peak.
The ``PS" and the ``VC" stand for pseudoscalar and vector channel,
respectively.
The ``stat." stands for statistical errors estimated by Jackknife method
and the ``sys." stands for systematic errors from SpM analyses in Appendix~\ref{sec:mu_determined}.
}
\label{tbl:spf_mean-err_LQCD}
\end{table*}
\renewcommand{\arraystretch}{1}

\begin{figure*}[tbp]
  \centering
  \begin{minipage}{.49\textwidth}
    \includegraphics[width=1.0\linewidth]{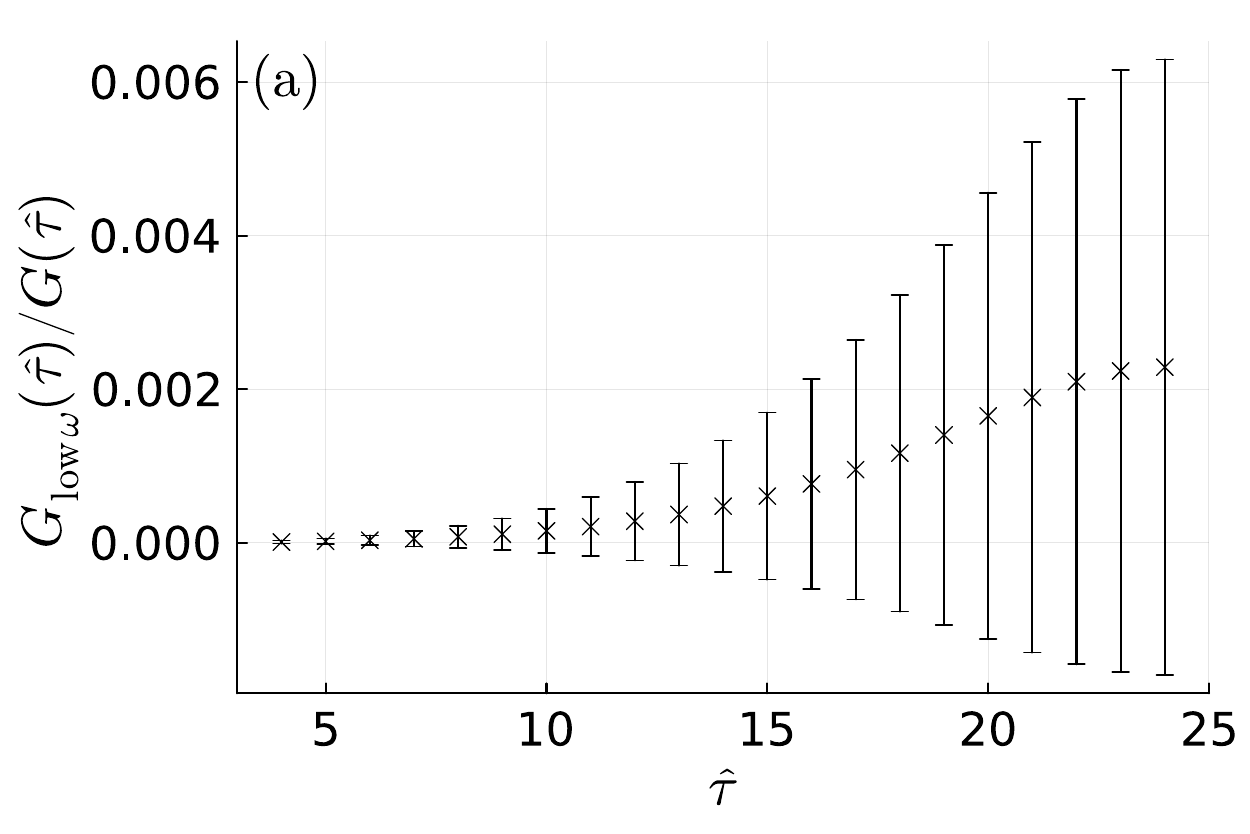}
  \end{minipage}
  \begin{minipage}{.49\textwidth}
    \includegraphics[width=1.0\linewidth]{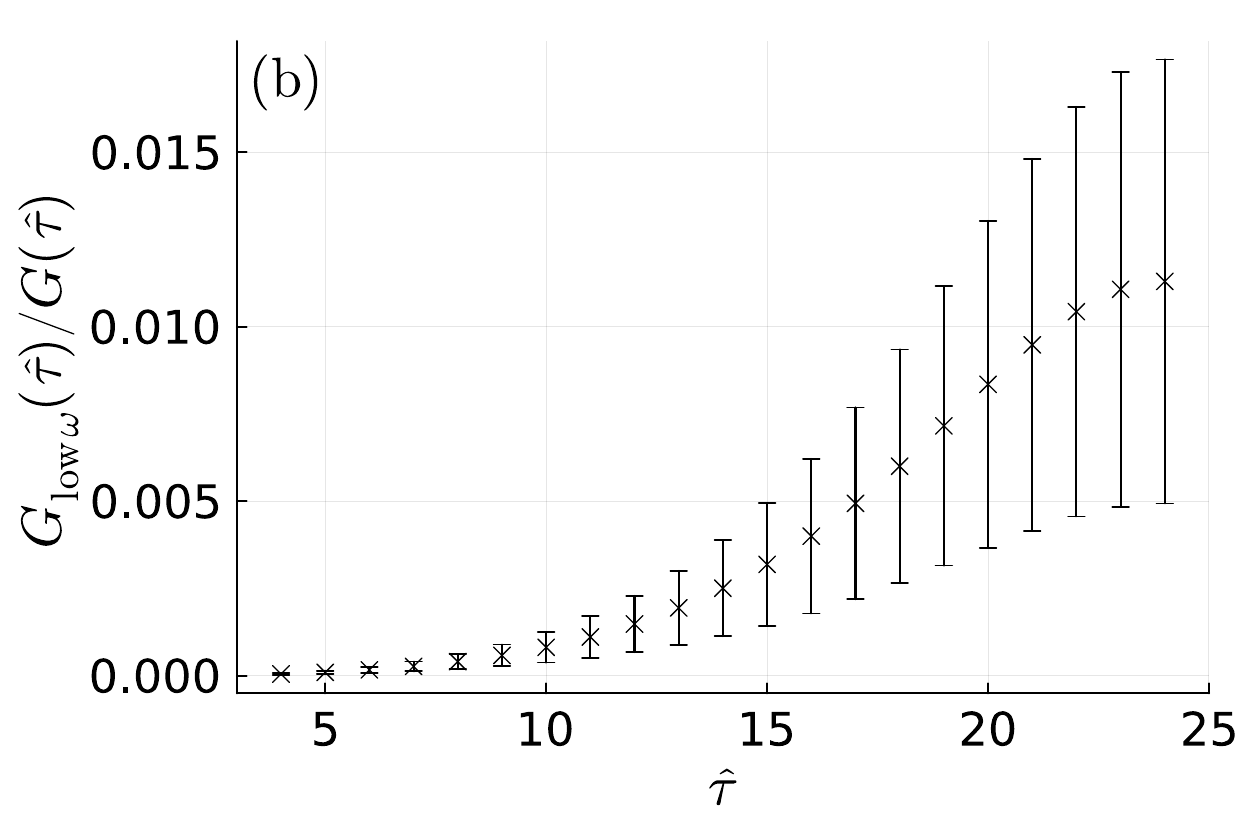}
  \end{minipage}
  \caption{
    Ratios $G_{\mathrm{low}\,\omega}(\hat{\tau})/G(\hat{\tau})$ of correlation functions calculated using the spectral functions shown in Fig.~\ref{fig:spf_LQCD_1.46Tc},
    where $G_{\mathrm{low}\,\omega}(\hat{\tau})$ is evaluated using the spectral function up to a frequency of 1.5 GeV
    and $G(\hat{\tau})$ over the entire frequency range.
    Figure (a) and (b) show the results in pseudoscalar and vector channels,
    respectively.
    The crosses denote the mean values of the ratios,
    and the vertical error bars represent the statistical errors of the ratios from Jackknife analyses.
  }
  \label{fig:ratio_correlator_LQCD_1.46Tc}
\end{figure*}

\indent
In low frequency region,
the value of the spectral function slightly fluctuates around $\rho=0$
and the statistical error of the spectral function
in the vicinity of $\omega=0$ is very large
in the pseudoscalar channel in Fig.~\ref{fig:spf_LQCD_1.46Tc}(a).
These behaviors are similar to those for $T<T_{\mathrm{c}}$
shown in Fig.~\ref{fig:spf_LQCD_0.73Tc}(a).
If these fluctuations in low-frequency region were meaningful, they should contribute to the correlation function.
In order to examine these contributions, we calculate the correlation function using the spectral function up to a frequency of 1.5 GeV.
Figure~\ref{fig:ratio_correlator_LQCD_1.46Tc}(a) shows the ratio $G_{\mathrm{low}\,\omega}(\hat{\tau})/G(\hat{\tau})$ of correlation functions calculated using the spectral functions shown in Fig.~\ref{fig:spf_LQCD_1.46Tc}(a),
where $G_{\mathrm{low}\,\omega}(\hat{\tau})$ is evaluated using the spectral function up to a frequency of 1.5 GeV
and $G(\hat{\tau})$ is evaluated over the entire frequency range.
The crosses denote the mean values of the ratios,
and the vertical error bars represent the statistical errors of the ratios from Jackknife analyses.
The results show it to be zero within the error range.
This finding suggests the absence of a transport peak.
\\
\indent
In the vector channel at $T>T_{\mathrm{c}}$,
although it is expected that the presence of the transport peak will be extracted,
in our analyses,
it does not appear as shown in Fig.~\ref{fig:spf_LQCD_1.46Tc}(b).
Furthermore,
the values of the spectral function in the vicinity of $\omega=0$ GeV are negative,
despite applying the constraint of positivity of the spectral function.
We confirm that
the positivity is rendered more protected as $\mu$ and $\mu^{\prime}$ increase,
and the values of the spectral function in the vicinity of $\omega=0$ GeV approach zero,
but do not turn to large positive values such as the transport peak.
In contrast to the pseudoscalar channel, however, the statistical error of the spectral function in the low-frequency region is small.
Therefore, the possibility of signal existence is considered.
As with the pseudoscalar channel, we calculate the correlation function using the spectral function up to a frequency of 1.5 GeV in order to examine its contribution to the correlation function.
Figure~\ref{fig:ratio_correlator_LQCD_1.46Tc}(b) shows the same result in Fig.~\ref{fig:ratio_correlator_LQCD_1.46Tc}(a) but for the vector channel.
This result indicates that,
in contrast to the case of pseudoscalar channel,
there is a statistically significant contribution.
This means that,
as seen in the mock-data test,
the contribution below the threshold appears to exist in the presence of the transport peak,
suggesting that some signal may indeed be present.
In order to estimate the transport with reasonable accuracy,
it may be necessary to make additional assumptions that extend beyond SpM,
including the shape of the transport peak.

\section{
Summary
\label{sec:summary}
}
We applied SpM to extract a spectral function
from a Euclidean-time meson correlation function. 
SpM is a method that solves inverse problems
by considering only the sparseness of the solution we seek.
Since the correlation function has correlation between different Euclidean times,
we took the covariance of the correlation function into account for SpM.
\\
\indent
First,
we tested SpM with mock data of the spectral function.
In order to check the applicability of SpM
for the number of data points of the correlation function
and the magnitude of errors of the correlation function,
we used the possible charmonium spectral function
for $T<T_{\mathrm{c}}$ and $T>T_{\mathrm{c}}$.
To examine whether the anticipated peak structure can be reproduced only when it is genuinely present, we also considered a free spectral function in which resonance peaks are intentionally absent.
This method was found to be poor at precisely reconstructing spectral functions with abrupt changes in value,
such as sharp peaks at $T<T_{\mathrm{c}}$ and transport peaks at $T>T_{\mathrm{c}}$.
However,
this test confirmed
that increasing the number of data points of the correlation function
and reducing the magnitude of errors of the correlation function
lead to output spectral functions closer to the input spectral function
and, when the input correlation function lacks information of a resonance peak,
this method does not generate spurious peak structures.
\\
\indent
Next,
we tried to extract the spectral functions from the charmonium correlation functions in pseudoscalar and vector channels
at $T\simeq 0.73T_{\mathrm{c}}$ with $N_{\tau}=96$ and $T\simeq 1.46T_{\mathrm{c}}$ with $N_{\tau}=48$ obtained by lattice QCD.
We got a spectral function with a broad peak around 4 GeV in each channel for $T<T_{\mathrm{c}}$,
which is slightly larger compared to the results in the previous study
using MEM~\cite{Ding2012.PhysRevD.86.014509}.
For $T\simeq 1.46T_{\mathrm{c}}$,
compared to the results for $T\simeq 0.73T_{\mathrm{c}}$,
we got a spectral function with a broader peak around 5 GeV in each channel,
which is also larger compared to the results from MEM.
In addition,
our calculation does not yield the transport peak,
which is expected to be the presence in the vector channel.
In order to estimate the transport peak,
it may be necessary to make assumptions that extend beyond the sparse modeling,
including the shape of the transport peak.
Although the positions and the widths of the resonance peak differ from the results from MEM
as described above,
the qualitative behavior is the same;
there is a physically consistent hierarchy in peak positions across channels,
and a clearly defined peak appears at lower temperatures that broadens at higher temperatures.
Therefore,
this result,
solely from the assumption of the sparse solution,
reflects underlying physics.

\begin{acknowledgments}
Our SpM analyses were performed
by a code created by E. Itou and Y. Nagai.
We are grateful to H.-T. Ding for sharing lattice data, and  E. Itou and Y. Nagai for fruitful discussions in the early stage of this work.
The work of A.T. was partially supported by JSPS KAKENHI Grant Numbers 20K14479, 22H05111, 22K03539 and JST BOOST, Japan Grant Number JPMJBY24F1. 
A.T. and H.O. were partially supported by JSPS KAKENHI Grant Number 22H05112.
This work was partially supported by MEXT as ``Program for Promoting Researches on the Supercomputer Fugaku''
(Grant Number JPMXP1020230411, JPMXP1020230409).
\end{acknowledgments}

\appendix

\section{
ADMM algorithm
\label{sec:ADMM}
}
In this appendix,
we briefly review the ADMM algorithm
following Refs.~\cite{itou2020sparse,Ohtsuki_JPSJ.89.012001}.
This algorithm,
which is developed by Ref.~\cite{Boyd.MAL-016},
solves the optimization problem that includes an $L_{1}$ regularization term with multiple constraints.
\\
\indent
We need to find a minimization of $F(\vec{\rho}^{\;\prime}_{W}\mid\vec{G}^{\prime}_{W},\lambda)$
in Eq.~\eqref{eq:cost_function_FW} with respect to $\vec{\rho}^{\;\prime}_{W}$
with two additional constraints in Eqs.~\eqref{eq:rho_positivity} and~\eqref{eq:rho_sum_rule}.
Following the conventional notation,
we change the variables as $\vec{\rho}\to\vec{x}$,
$\vec{\rho}^{\;\prime}_{W}\to\vec{x}^{\prime}$,
$\vec{G}_{W}\to\vec{y}$,
and $\vec{G}^{\prime}_{W}\to\vec{y}^{\prime}$,
and the matrix as $S_{W}\to S$.
Then the cost function is rewritten as
\begin{equation}
    F(\vec{x}^{\prime}|\vec{y}^{\prime},\lambda)
    =\frac{1}{2}||\vec{y}^{\prime}-S\vec{x}^{\prime}||^{2}_{2}+\lambda||\vec{x}^{\prime}||_{1},
\end{equation}
and the constraints are represented as
\begin{equation}
    x_{j}\ge 0,
    \quad
    \sum_{j}x_{j}=1.
\end{equation}
The dimension of this optimization problem is given by
$L=\mathrm{min}(M,N)$,
where $M$ and $N$ are the sizes of $\vec{y}$ and $\vec{x}$,
respectively.
In fact,
we can reduce $L$ by introducing a cut of the singular value in analysis,
and we mentioned the method of reducing the dimension in Sec.~\ref{subsec:procedure_SpM}.
\\
\indent
In the ADMM algorithm,
we introduce auxiliary vectors $\vec{z}$ and $\vec{z}^{\prime}$,
and consider minimization of the function
\begin{align}
    \tilde{F}(\vec{x}^{\prime},\vec{z}^{\prime},\vec{z})
    =\frac{1}{2\lambda}||\vec{y}^{\prime}-S\vec{x}^{\prime}||^{2}_{2}
    -\nu(\langle V\vec{x}^{\prime}\rangle-1)\nonumber\\
    +||\vec{z}^{\prime}||_{1}
    +\lim_{\gamma\to\infty}\gamma\sum_{j}\Theta(-z_{j}),
\end{align}
subject to
\begin{equation}
    \vec{z}^{\prime}=\vec{x}^{\prime},
    \quad
    \vec{z}=V\vec{x}^{\prime},
    \label{eq:z'=x',z=Vx'}
\end{equation}
where we have used a convention that primed vectors denote
quantities represented in the IR basis,
$V$ is the matrix obtained by the SVD,
$\langle V\vec{x}^{\prime}\rangle$ represents the sum of the components of $V\vec{x}^{\prime}$,
and $\Theta$ is the Heaviside step function.
The sum rule is imposed by the Lagrange multiplier $\nu$,
and non-negativity is expressed by an infinite potential with $\gamma$.
The auxiliary vectors $\vec{z}$ and $\vec{z}^{\prime}$
are in charge of the $L_{1}$ regularization and the non-negativity,
respectively.
\\
\indent
The constraints for the auxiliary vectors,
which are given in Eq.~\eqref{eq:z'=x',z=Vx'},
are treated by the augmented Lagrange multiplier method.
In order to treat the first constraint,
$\vec{z}^{\prime}=\vec{x}^{\prime}$,
normalized Lagrange multipliers $\vec{u}^{\prime}$
and a parameter $\mu^{\prime}$ are introduced.
Similarly,
in order to treat the second constraint,
$\vec{z}=V\vec{x}^{\prime}$,
normalized Lagrange multipliers $\vec{u}$
and a parameter $\mu$ are introduced.
With the introduction of $\vec{u}$, $\vec{u}^{\prime}$, $\mu$ and $\mu^{\prime}$,
the loss function actually applied to the ADMM algorithm is expressed as
\begin{align}
    \tilde{L}_{\mathrm{aug.}}
    =&\frac{1}{2\lambda}||\vec{y}^{\prime}-S\vec{x}^{\prime}||^{2}_{2}
    +||\vec{z}^{\prime}||_{1}\nonumber\\
    &+\frac{\mu^{\prime}}{2}||\vec{x}^{\prime}-\vec{z}^{\prime}+\vec{u}^{\prime}||^{2}_{2}
    -\frac{\mu^{\prime}}{2}||\vec{u}^{\prime}||^{2}_{2}\nonumber\\
    &+\frac{\mu}{2}||V\vec{x}^{\prime}-\vec{z}+\vec{u}||^{2}_{2}
    -\frac{\mu}{2}||\vec{u}||^{2}_{2}\nonumber\\
    &-\nu(\langle V\vec{x}^{\prime}\rangle-1)
    +\lim_{\gamma\to\infty}\gamma\sum_{j}\Theta(-z_{j}).
\end{align}
While $\vec{u}$ and $\vec{u}^{\prime}$ are iteratively updated
together with its conjugate vectors $\vec{z}$ and $\vec{z}^{\prime}$,
respectively,
the coefficients $\mu$ and $\mu^{\prime}$ are predetermined.
In addition,
$\mu$ and $\mu^{\prime}$ control speed of convergence.
\\
\indent
Considering the search for the minimum value of the loss function $\tilde{L}_{\mathrm{aug.}}$,
we obtain update formulas used in actual computations:
\begin{align}
    \vec{x}^{\prime}\leftarrow\;
    &\left(
        \frac{1}{\lambda}S^{\mathrm{t}}S
        +(\mu^{\prime}+\mu)I
     \right)^{-1}\nonumber\\
    &\times\left(
        \frac{1}{\lambda}S^{\mathrm{t}}\vec{y}^{\prime}
        +\mu^{\prime}(\vec{z}^{\prime}-\vec{u}^{\prime})
        +\mu V^{\mathrm{t}}(\vec{z}-\vec{u})
        +\nu V^{\mathrm{t}}\vec{e}
     \right)\nonumber\\
    &\equiv\vec{\xi}_{1}+\nu\vec{\xi}_{2},
    \label{eq:x'_update}\\
     \vec{z}^{\prime}\leftarrow\;
    &\mathcal{S}_{1/\mu^{\prime}}(\vec{x}^{\prime}+\vec{u}^{\prime}),\\
     \vec{u}^{\prime}\leftarrow\; 
    &\vec{u}^{\prime}+\vec{x}^{\prime}-\vec{z}^{\prime},\\
     \vec{z}\leftarrow\;
    &\mathcal{P}_{+}(V\vec{x}^{\prime}+\vec{u}),
    \label{eq:z_update}\\
     \vec{u}\leftarrow\;
    &\vec{u}+V\vec{x}^{\prime}-\vec{z},
    \label{eq:u_update}
\end{align}
where $I$ is the unit matrix,
$e_{i}=1$,
\begin{equation}
    \nu=\frac{1-\langle V\vec{\xi}_{1}\rangle}{\langle V\vec{\xi}_{2}\rangle}.
\end{equation}
Here,
$\mathcal{P}_{+}$ denotes a projection operator onto non-negative quadrant,
i.e.,
$\mathcal{P}_{+}(z_{j})=\mathrm{max}(z_{j},0)$ for each element,
and $\mathcal{S}_{\alpha}$ denotes the element-wise soft thresholding function,
which is defined for each element by
\begin{equation}
    \mathcal{S}_{\alpha}(x)=\left\{
        \begin{array}{ll}
           x-\alpha & (x>\alpha) \\
           0 & (-\alpha\le x\le\alpha)\\
           x+\alpha & (x<-\alpha)
        \end{array}
    \right.
    .
\end{equation}
The updates in Eqs.~\eqref{eq:x'_update}--\eqref{eq:u_update}
are iteratively performed until convergence is reached.
\\
\indent
In our SpM analyses,
whether convergence has been achieved is determined
by looking at the value of the residuals,
$||\vec{z}-V\vec{x}^{\prime}||_{1}/N_{\omega}$.
Also,
$\nu$ is set to zero since the condition of the sum rule is not applied,
the initial vectors are set at zero vectors for all
and we set the coefficients $\mu$ and $\mu^{\prime}$ at $10^2-10^4$.
The actual values of $\mu$ and $\mu^{\prime}$ are written in Sects.~\ref{sec:mock_data_tests} and \ref{sec:results_LQCD},
and we explain how to determine these values of $\mu$ and $\mu^{\prime}$
in Appendix~\ref{sec:mu_determined}.

\section{
Singular values of the kernel $K_{W}$
\label{sec:SW_of_KW}
}
\begin{figure}[htbp]
  \centering
    \includegraphics[width=1.0\linewidth]{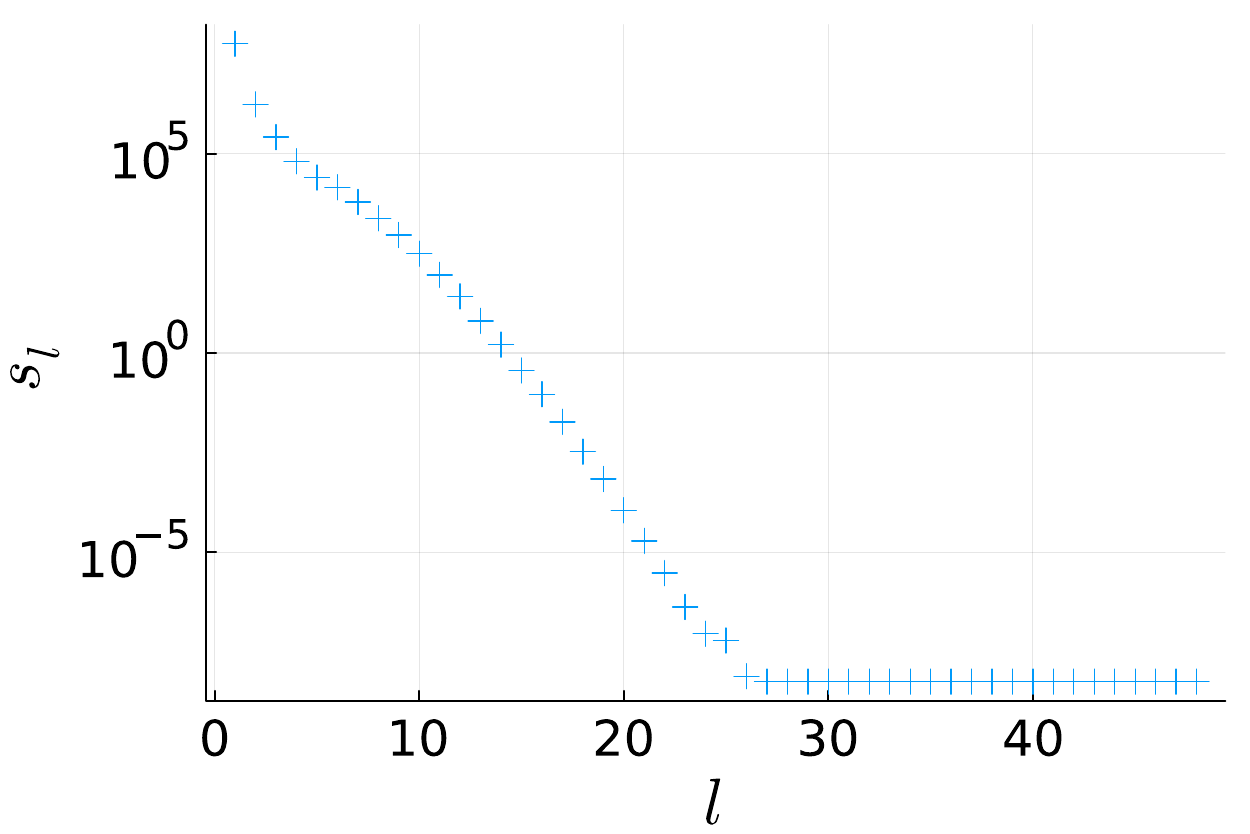}
  \caption{
    Singular values of the kernel $K_{W}^{(0)}$
    with $N_{\tau}=96$ and $\varepsilon=5\times 10^{-3}$.
  }
  \label{fig:sv_rho}
\end{figure}

One of the key to solving the inverse problem by SpM is dimensionality reduction.
In order to utilize this reduction,
it is necessary that
the components of the singular values of the kernel decrease exponentially.
In previous studies~\cite{itou2020sparse,Otsuki.PhysRevE.95.061302},
kernels of the type given by Eq.~\eqref{eq:1_kernel} have been used,
and their singular values have been shown to decrease exponentially.
On the other hand,
our study employs the kernel $K^{(n)}_{W}$ defined by Eq.~\eqref{eq:kernel_W}.
The singular value of this kernel must be investigated.
\\
\indent
Figure~\ref{fig:sv_rho} shows the singular values of the kernel $K_{W}^{(0)}$
given in Eqs.~\eqref{eq:K_tau_hat_omega_hat} and~\eqref{eq:kernel_W}
with $N_{\tau}=96$ and $\varepsilon=5\times 10^{-3}$.
The singular values decrease exponentially,
indicating that dimensionality reduction can be used.
It has also been made sure that
the singular values of the kernel $K_{W}^{(0)}$ decrease exponentially
for other combinations of $N_{\tau}$ and $\varepsilon$ as well.

\begin{figure}[htbp]
  \centering
    \includegraphics[width=1.0\linewidth]{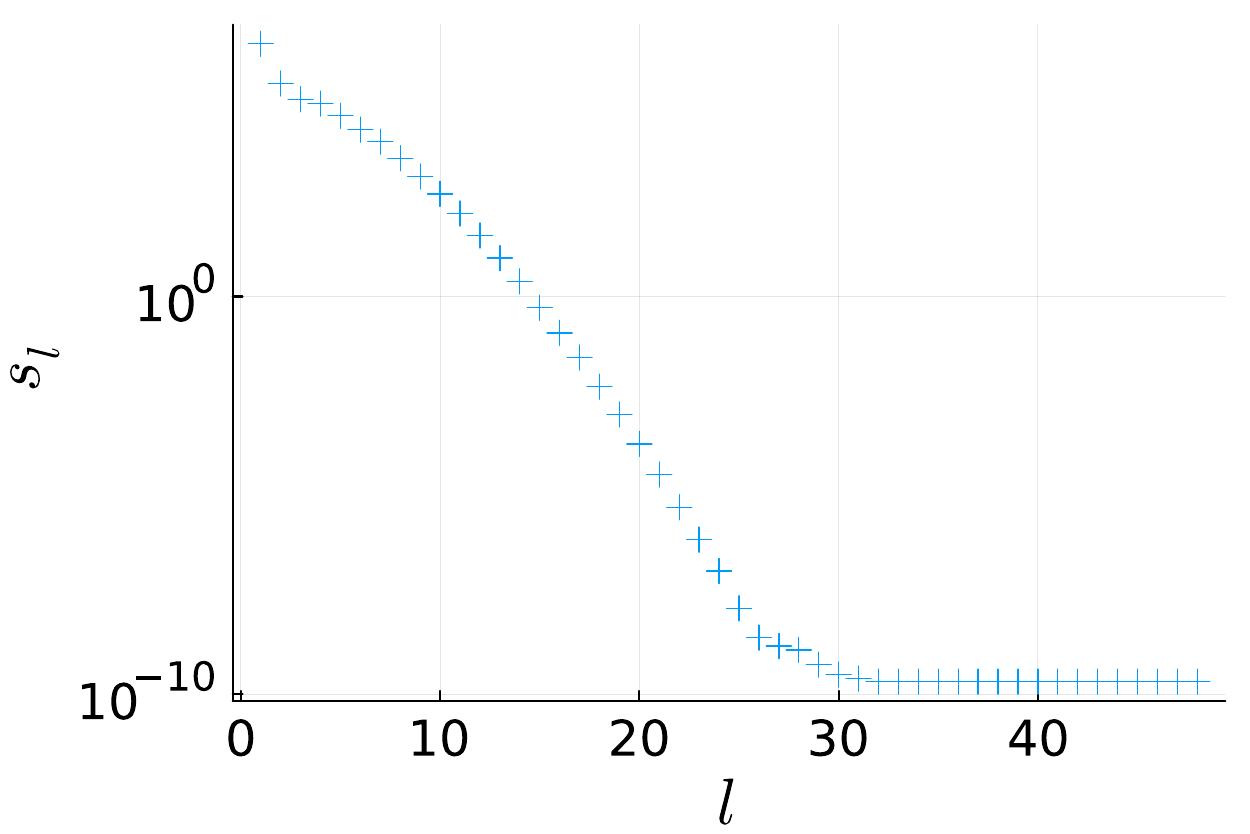}
  \caption{
    Same as Fig.~\ref{fig:sv_rho} but for the kernel $K_{W}^{(1)}$.
  }
  \label{fig:sv_rho-omega}
\end{figure}

Figure~\ref{fig:sv_rho-omega} shows the same as in Fig.~\ref{fig:sv_rho}
but for the kernel $K_{W}^{(1)}$
given in Eqs.~\eqref{eq:K1_tau_hat_omega_hat} and~\eqref{eq:kernel_W}
with $N_{\tau}=96$ and $\varepsilon=5\times 10^{-3}$.
As in the case of $K_{W}^{(0)}$,
the singular values decrease exponentially,
indicating that dimensionality reduction can be used.
It has also been made sure that
the singular values of the kernel $K_{W}^{(1)}$ decrease exponentially
for other combinations of $N_{\tau}$ and $\varepsilon$ as well.

\section{
Detailed formalism for extracting spectral functions in SpM
\label{sec:formalism_ap}
}
Here we explain a detailed formalism
for extracting spectral functions in SpM.
\\
\indent
The starting points of the formalism are
Eqs.~\eqref{eq:spf_Lehmann_rep_0} and~\eqref{eq:1_kernel}.
Namely,
the correlation function $G(\tau)$ is written as
\begin{equation}
    \displaystyle G(\tau)=\int^{\infty}_{0}d\omega
    \frac{\cosh\left[
      \omega\left(
      \tau-\frac{1}{2T}
      \right)
      \right]}{\sinh\left(
    \frac{\omega}{2T}
    \right)}\rho(\omega).
\end{equation}
To treat it on a lattice whose spacing is $a$,
it can be written in terms of dimensionless quantities as
\begin{equation}
    \displaystyle \hat{G}(\hat{\tau})
    =\int^{\infty}_{0}d\hat{\omega}
    \frac{\cosh\left[
      \hat{\omega}\left(
      \hat{\tau}-N_{\tau}/2
      \right)
      \right]}{\sinh\left(
    \hat{\omega}N_{\tau}/2
    \right)}\hat{\rho}(\hat{\omega}),
    \label{eq:Ghat(tauhat)}
\end{equation}
where $\hat{\omega}\equiv\omega a$,
$\hat{\tau}\equiv\tau/a$,
$\hat{G}\equiv Ga^{3}$,
$\hat{\rho}\equiv\rho a^{2}$
and $N_{\tau}$ is the extent of Euclidean times,
respectively.
As can be seen from the meaning of $N_{\tau}$,
$N_{\tau}a = 1/T$.
Now we introduce $\hat{\omega}_{\mathrm{max}}\equiv\omega_{\mathrm{max}}a$
as a cutoff for $\hat{\omega}$
and normalize $\hat{\omega}$ and $\hat{\tau}$
by $\hat{\omega}_{\mathrm{max}}$ and $N_{\tau}$ respectively as follows:
\begin{equation}
    \omega^{\prime}\equiv\frac{\hat{\omega}}{\hat{\omega}_{\mathrm{max}}},
    \quad
    \tau^{\prime}\equiv\frac{\hat{\tau}}{N_{\tau}}.
\end{equation}
Clearly,
the ranges of these primed variables are $0\le\omega^{\prime}\le 1$ and $0\le\tau^{\prime}<1$.
Thus,
with these primed variables,
Eq.~\eqref{eq:Ghat(tauhat)} is written as
\begin{equation}
    \displaystyle \hat{G}(\tau^{\prime})
    =\int^{1}_{0}d\omega^{\prime}\hat{\omega}_{\mathrm{max}}
    \frac{\cosh\left[
      \omega^{\prime}\Lambda\left(
      \tau^{\prime}-1/2
      \right)
      \right]}{\sinh\left(
    \omega^{\prime}\Lambda/2
    \right)}\hat{\rho}(\omega^{\prime}),
    \label{eq:Ghat(tau')}
\end{equation}
where $\Lambda\equiv\hat{\omega}_{\mathrm{max}}N_{\tau}$.
By discretizing the definite integral,
turning it into a finite sum,
and replacing $d\omega^{\prime}$ with $\Delta\omega^{\prime}$,
the finite sum is given as
\begin{equation}
   \displaystyle \hat{G}(\tau^{\prime}_{i})
   =\sum^{N_{\omega}-1}_{j=0}\Delta\omega^{\prime}
   \hat{\omega}_{\mathrm{max}}
   \frac{\cosh\left[
      \omega^{\prime}_{j}\Lambda\left(
      \tau^{\prime}_{i}-1/2
      \right)
      \right]}{\sinh\left(
    \omega^{\prime}_{j}/2
    \right)}\hat{\rho}(\omega^{\prime}_{j}),
    \label{eq:Ghat(tau'_i)}
\end{equation}
where $\Delta\omega^{\prime}\equiv 1/(N_{\omega}-1)$
and $N_{\omega}$ is the number of points in $\omega$-space.
Note that $\omega^{\prime}_{0}=0$ and $\tau^{\prime}_{0}=0$.
\\
\indent
Here we address two inconveniences.
One is the lattice cutoff effects
which appear correlation functions at short distances.
To reduce it,
we remove the first few points of the input data.
The first point of Euclidean time after the removal is defined
as $\tau^{\prime}_{\mathrm{r}}$ as the reference Euclidean time.
The other is
that the right-hand side of Eq.~\eqref{eq:Ghat(tau'_i)}
diverges at $\omega^{\prime}_{0}=0$.
To deal with it,
we define
\begin{equation}
    \tilde{\rho}(\omega^{\prime}_{j})
    \equiv\frac{\cosh\left[
      \omega^{\prime}_{j}\Lambda\left(
      \tau^{\prime}_{\mathrm{r}}-1/2
      \right)
      \right]}{\sinh\left(
    \omega^{\prime}_{j}\Lambda/2
    \right)}\hat{\rho}(\omega^{\prime}_{j}).
    \label{eq:rho_tilde}
\end{equation}
Rewriting Eq.~\eqref{eq:Ghat(tau'_i)} using $\tilde{\rho}(\omega^{\prime}_{j})$,
we obtain
\begin{equation}
    \displaystyle \hat{G}(\tau^{\prime}_{i})
   =\sum^{N_{\omega}-1}_{j=0}\Delta\omega^{\prime}
   \hat{\omega}_{\mathrm{max}}
   \frac{\cosh\left[
      \omega^{\prime}_{j}\Lambda\left(
      \tau^{\prime}_{i}-1/2
      \right)
      \right]}{\cosh\left[
      \omega^{\prime}_{j}\Lambda\left(
      \tau^{\prime}_{\mathrm{r}}-1/2
      \right)
      \right]}\tilde{\rho}(\omega^{\prime}_{j}).
    \label{eq:Ghat(tau'_i)/Ghat(0)_rhotilde}
\end{equation}
\\
\indent
In Ref.~\cite{itou2020sparse},
$\sqrt{\Delta\omega^{\prime}}$ is included in the kernel
to make the dependence of $N_{\omega}$ milder in the SpM analysis.
We adopt it too and define the kernel as
\begin{equation}
    K(\tau^{\prime}_{i},\omega^{\prime}_{j})
    \equiv\sqrt{\Delta\omega^{\prime}}
   \hat{\omega}_{\mathrm{max}}
   \frac{\cosh\left[
      \omega^{\prime}_{j}\Lambda\left(
      \tau^{\prime}_{i}-1/2
      \right)
      \right]}{\cosh\left[
      \omega^{\prime}_{j}\Lambda\left(
      \tau^{\prime}_{\mathrm{r}}-1/2
      \right)
      \right]}.
      \label{eq:discretized_kernel}
\end{equation}
Rewriting the primed variables in Eq.~\eqref{eq:discretized_kernel}
into hatted variables yields Eq.~\eqref{eq:K_tau_hat_omega_hat}.

\section{
Estimation of an optimal value of $\lambda$
\label{sec:lambda_opt}
}
\begin{figure*}[htb]
  \centering
  \begin{minipage}{.49\textwidth}
    \includegraphics[width=1.0\linewidth]{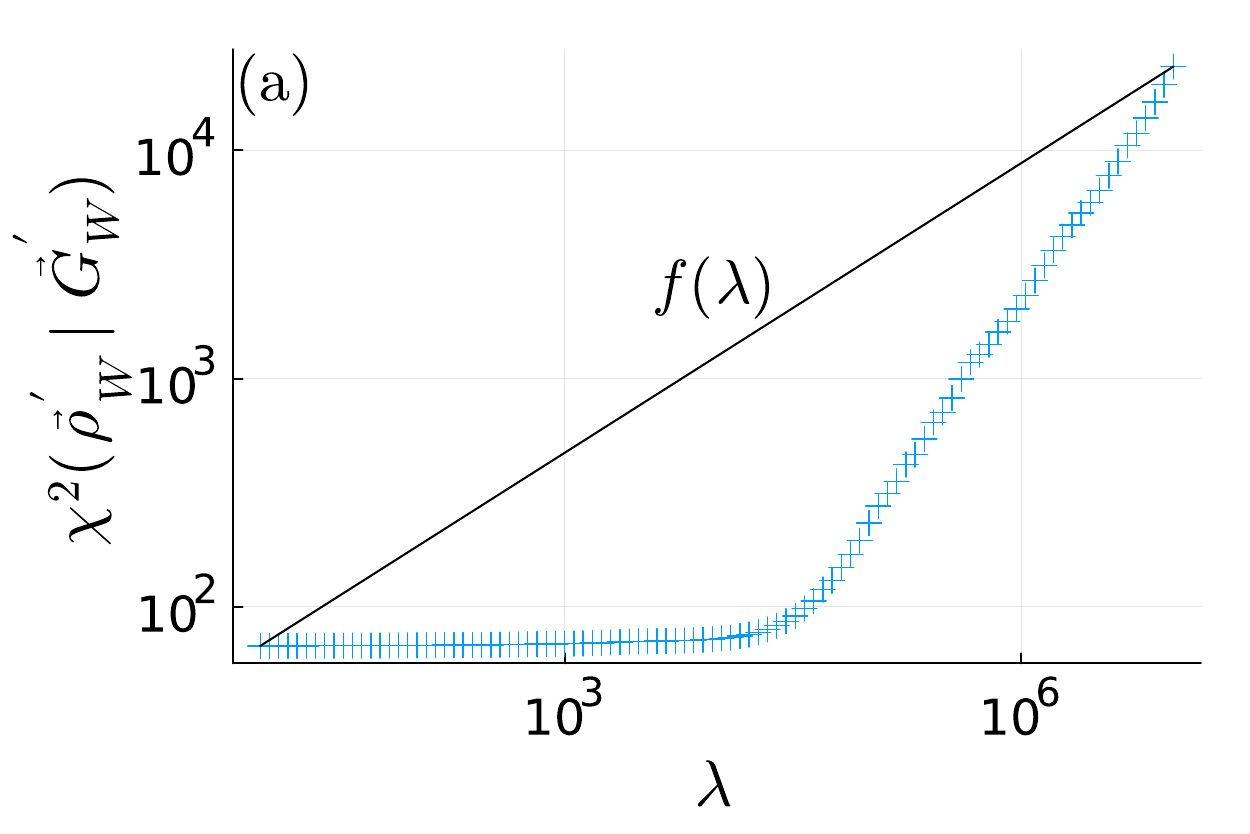}
  \end{minipage}
  \begin{minipage}{.49\textwidth}
    \includegraphics[width=1.0\linewidth]{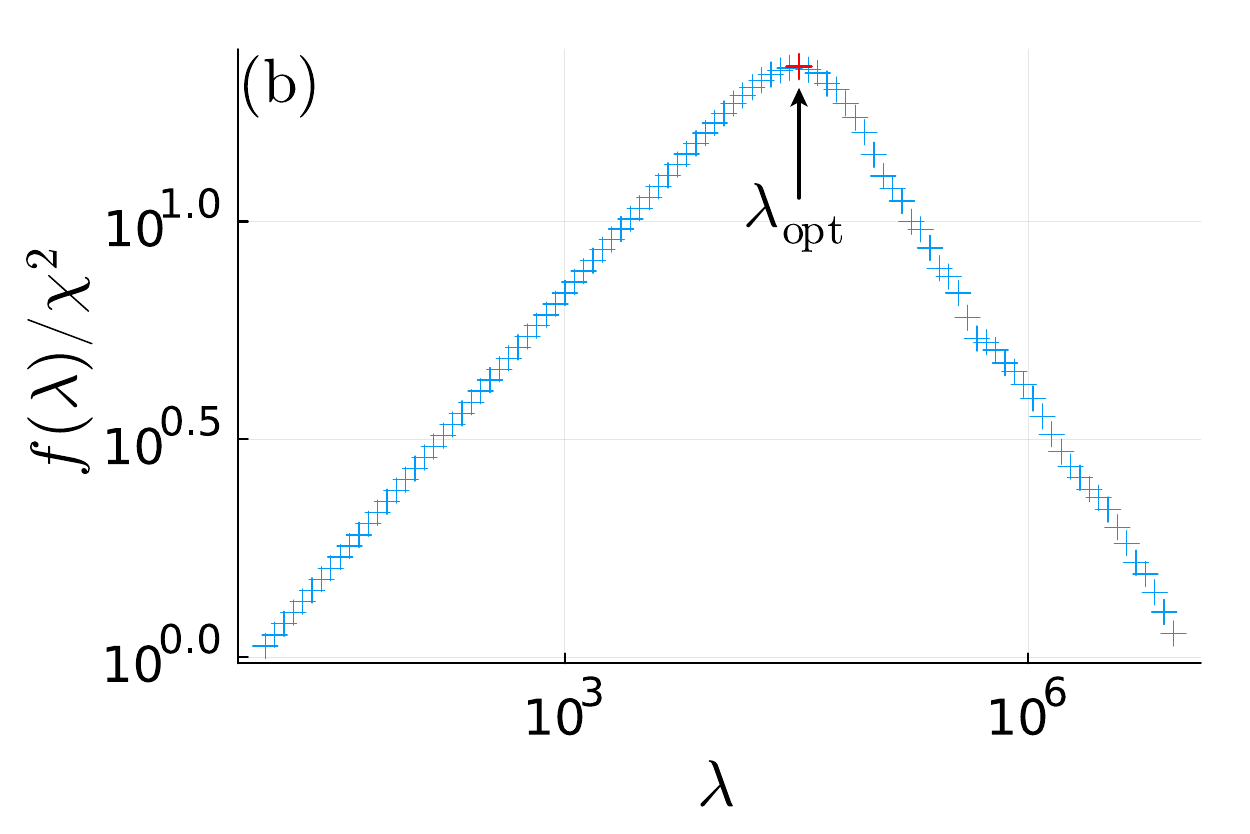}
  \end{minipage}
  \caption{
    An example of square error and $f(\lambda)/\chi^{2}$
    in the mock-data test for $T>T_{\mathrm{c}}$ with $N_{\tau}=64$ and $\varepsilon=10^{-5}$.
    (a) The square error in IR basis,
    $\chi^{2}(\rho^{\prime}_{W}(\hat{\omega}_{j})\mid G^{\prime}_{W}(\hat{\tau}_{i}))$,
    as a function of $\lambda$.
    The function $f(\lambda)$ which is represented by the black solid line
    is defined as a linear function
    that connects the value of $\chi^{2}$ at $\lambda_{\mathrm{min}}$
    and that at $\lambda_{\mathrm{max}}$.
    This connection is expressed on a logarithmic scale.
    (b) $f(\lambda)/\chi^{2}$ as a function of $\lambda$,
    which means a reduction of $\chi^{2}$ relative to $f(\lambda)$.
    The red plus sign denotes the peak of $f(\lambda)/\chi^{2}$
    and represents the optimal value of $\lambda$,
    $\lambda_{\mathrm{opt}}$.
  }
  \label{fig:lambda-vs-chi2_fchi}
\end{figure*}

In this appendix
we explain the estimation of an optimal value of $\lambda$,
$\lambda_{\mathrm{opt}}$.
We employ the same method as in the previous study~\cite{itou2020sparse},
which is summarized as follows:
\begin{enumerate}
    \item Fix a search range of $\lambda$
    as $[\lambda_{\mathrm{min}}, \lambda_{\mathrm{max}}]$
    and divide the range into equal parts $N_{\lambda}$.
    \item Solve optimization problems using the squared error in IR basis,
    $\chi^{2}(\rho^{\prime}_{W}(\hat{\omega}_{j})\mid G^{\prime}_{W}(\hat{\tau}_{i}))$,
    with various values of $\lambda$ in the fixed range,
    $[\lambda_{\mathrm{min}}, \lambda_{\mathrm{max}}]$,
    by using ADMM algorithm~\cite{Boyd.MAL-016}
    with the positivity constraint $\hat{\rho}(\omega^{\prime}_{i})\ge 0$.
    \item Connect the value of $\chi^{2}$ at $\lambda_{\mathrm{min}}$
    with that at $\lambda_{\mathrm{max}}$ 
    by using a linear function on a logarithmic scale, $f(\lambda)$,
    since we obtain $\chi^2(\rho^{\prime}_{W}(\hat{\omega}_{j})\mid G^{\prime}_{W}(\hat{\tau}_{i}))$
    as a function of $\lambda$.
    \item Obtain $\lambda_{\mathrm{opt}}$ which produces the peak of $f(\lambda)/\chi^{2}$.
\end{enumerate}
In our SpM analyses,
$N_{\lambda}$ is set at 100,
and the range of $[\lambda_{\mathrm{min}}, \lambda_{\mathrm{max}}]$
is set to $[0.1,10^7]$ for mock-data tests and $[1.0,10^7]$ for analyses of lattice QCD data.
\\
\indent
The $\lambda_{\mathrm{opt}}$ corresponds to the kink position in $\chi^{2}$,
and this position is the best place
for the parameter $\lambda$ of SpM~\cite{Otsuki.PhysRevE.95.061302}.
Figure~\ref{fig:lambda-vs-chi2_fchi} shows
an example of square error and $f(\lambda)/\chi^{2}$ in our mock-data test
for $T>T_{\mathrm{c}}$ with $N_{\tau}=64$ and $\varepsilon=10^{-5}$.
Figure~\ref{fig:lambda-vs-chi2_fchi} (a) shows the square error in IR basis,
$\chi^{2}(\rho^{\prime}_{W}(\hat{\omega}_{j})|G^{\prime}_{W}(\hat{\tau}_{i}))$,
as a function of $\lambda$.
The black solid line represents $f(\lambda)$.
The purpose of the function $f(\lambda)$ is to accurately identify
the kink position of $\chi^{2}$ as a function of $\lambda$.
Using this $f(\lambda)$,
the peak position of $f(\lambda)/\chi^{2}$ corresponds to
the kink position of $\chi^{2}$,
as shown in Fig.~\ref{fig:lambda-vs-chi2_fchi} (b).
In Fig.~\ref{fig:lambda-vs-chi2_fchi} (b),
the red plus sign denotes the peak of $f(\lambda)/\chi^{2}$
and represents the optimal value of $\lambda$,
$\lambda_{\mathrm{opt}}$.

\indent
In our mock-data tests,
$\lambda_{\mathrm{opt}}$ is estimated
using the average value of 300 correlation functions
created with Gaussian random numbers as error.
These 300 correlation functions are created
only for the purpose of calculating the covariance matrix.
The statistical error of the spectral functions is not calculated
in the mock-data tests for simplicity.
Thus,
there is only one $\lambda_{\mathrm{opt}}$
to be estimated for each combination of $\varepsilon$ and $N_{\tau}$.
On the other hand,
in the analyses of actual lattice QCD data,
the statistical error of the spectral function is calculated using the Jackknife method.
In this method,
Jackknife samples are produced.
While it is possible to determine $\lambda_{\mathrm{opt}}$ for each Jackknife sample,
the calculation of $\lambda_{\mathrm{opt}}$ is time-consuming.
Furthermore,
when an attempt was made to calculate $\lambda_{\mathrm{opt}}$
for each jackknife sample at a single temperature and for a single channel,
it was found that the resulting $\lambda_{\mathrm{opt}}$ values were within the same order of magnitude, 
and the differences in the spectral function due to variations in $\lambda_{\mathrm{opt}}$ were found to be negligible.
Consequently,
as in the mock-data test,
the mean value of the correlation function is utilized to calculate $\lambda_{\mathrm{opt}}$.
This $\lambda_{\mathrm{opt}}$ is then used for each Jackknife sample.
Thus,
there is only one $\lambda_{\mathrm{opt}}$
to be estimated for each combination of a temperature and a channel.

\section{
Determination of $\mu$ and $\mu^{\prime}$
\label{sec:mu_determined}
}
\begin{figure*}[htbp]
  \centering
  \includegraphics[width=0.6\linewidth]{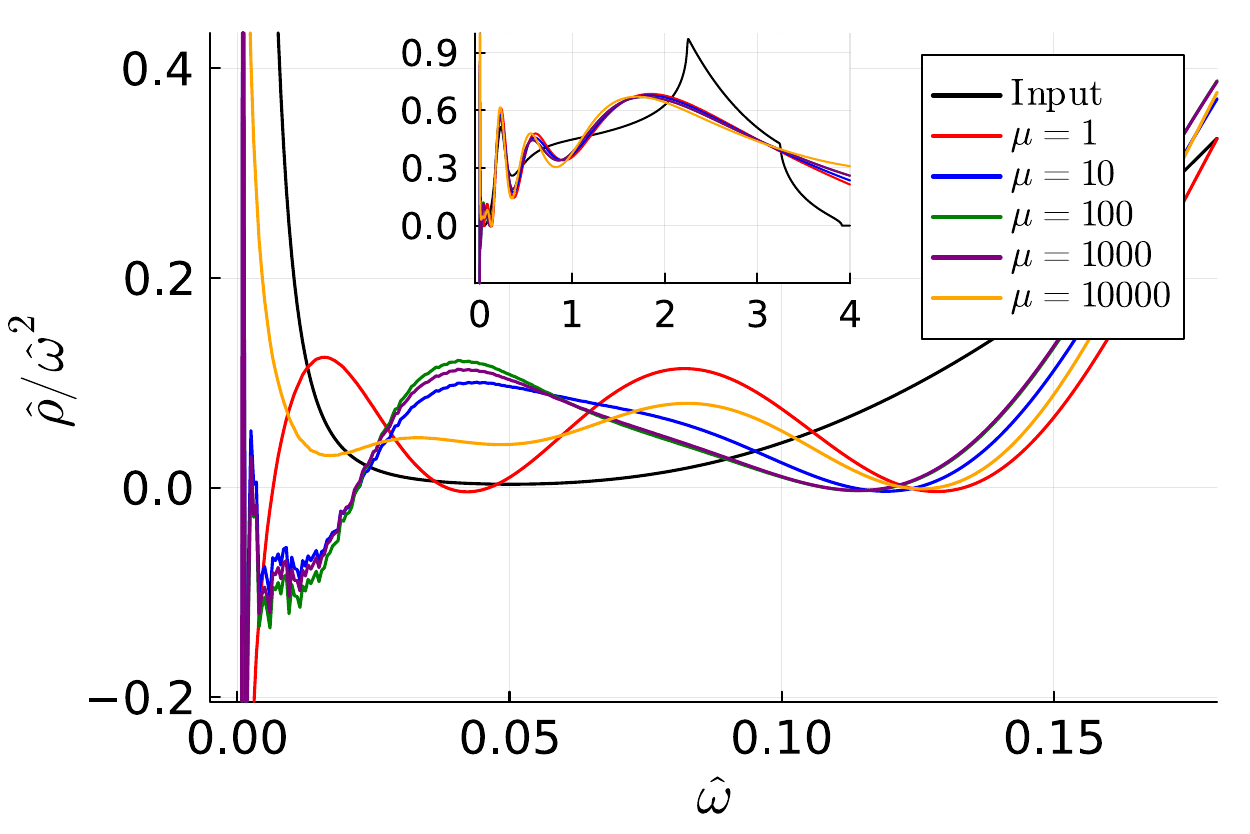}
  \caption{
    Spectral functions in small value of $\hat{\omega}$ at various values of $\mu$,
    which are obtained by using $K^{(0)}_{W}(\hat{\tau}_{i},\hat{\omega}_{j})$
    in the mock-data test for $T>T_{\mathrm{c}}$ with $N_{\tau}=96$ and $\varepsilon=10^{-2}$.
    The black line represents the input spectral function,
    and the red, blue, green, purple and yellow lines represent
    output results with $\mu=\mu^{\prime}=1$, 10, 100, 1000, 10000,
    respectively.
    These spectral functions in whole $\hat{\omega}$ region are drawn in small plot inside.  
  }
  \label{fig:spf_mu_comp_zoom_p-in-p}
\end{figure*}

The parameter $\mu$ is introduced to impose the constraint,
$\vec{z}=V\vec{x}^{\prime}$.
As can be seen in the ADMM algorithm,
the larger the magnitude of $\mu$,
the more vectors are updated to respect that constraint.
Also,
as expressed in Eqs.~\eqref{eq:z_update},
$\vec{z}$ is always positive definite.
Therefore,
satisfying $\vec{z}=V\vec{x}^{\prime}$
is almost equivalent to satisfying the positivity of the spectral function $\vec{x}$
that we want to find.
\\
\indent
It is not obvious what value of $\mu$ should be set to.
Since we need to focus on the number of input data $N_{\tau}$
and the magnitude of noise $\varepsilon$
in order to investigate the applicability of SpM in mock-data tests,
we want to fix the value of $\mu$.
To this end,
we have sought the optimal $\mu$ by calculating spectral functions for various values of $\mu$.
For simplicity,
$\mu$ and $\mu^{\prime}$ have been set to the same value in this study,
and the following values for $\mu$ and $\mu^{\prime}$ have been investigated:
$\mu=\mu^{\prime}=10^{i}\;(i=0,\,1,\,2,\,3,\,4,\,5)$.
\\
\indent
As an example,
we draw the spectral function at various values of $\mu$ in Fig.~\ref{fig:spf_mu_comp_zoom_p-in-p}.
Figure~\ref{fig:spf_mu_comp_zoom_p-in-p} shows
spectral functions in small value of $\hat{\omega}$
obtained by using $K^{(0)}_{W}(\hat{\tau}_{i},\hat{\omega}_{j})$
in the mock-data test for $T>T_{\mathrm{c}}$ with $N_{\tau}=96$ and $\varepsilon=10^{-2}$.
The black line represents the input spectral function,
and the red, blue, green, purple and yellow lines represent
output results with $\mu=\mu^{\prime}=1$, 10, 100, 1000, 10000,
respectively.
These spectral functions in whole $\hat{\omega}$ region are drawn in small plot inside.
When $\mu = 10^4$,
it can be seen that the spectral function always has a positive value,
while for other values,
there are sections where it has a negative value.
It is confirmed that the spectral function always has a positive value even when $\mu = 10^5$.
However,
when the value of $\mu$ is excessively large,
a discernible peak in $f(\lambda)/\lambda$ is not evident,
unlike Figure~\ref{fig:lambda-vs-chi2_fchi}(b).
Consequently,
it is not possible to estimate $\lambda_{\mathrm{opt}}$ appropriately.
Thus it is determined that
$\mu=10^{5}$ is not suitable option.

\indent
The same investigation is conducted for all combinations of kernel $K^{(n)}_{W}$, $N_{\tau}$, $\varepsilon$, and temperature $T$
used in the mock-data test.
As a result,
it is determined that
$\mu=\mu^{\prime}=10^{4}$ is optimal for both temperatures $T<T_{\mathrm{c}}$ and $T>T_{\mathrm{c}}$
when using $K^{(0)}_{W}$,
and $\mu=\mu^{\prime}=10^{3}$ is optimal for $T<T_{\mathrm{c}}$ when using $K^{(1)}_{W}$.
\\
\indent
A thorough investigation is conducted using actual lattice QCD data,
with $\mu=\mu^{\prime}=10^{i}\;(i=0,\,1,\,2,\,3,\,4)$
in order to ascertain the appropriate value of $\mu$.
Consequently,
it is ascertained that the $\lambda_{\mathrm{opt}}$ is appropriately obtained for $\mu=\mu^{\prime}=10^i\;(i=2,\,3,\,4)$,
and ensuring the preservation of the positivity of spectral function.
Instead of setting the value of $\mu$ or $\mu^{\prime}$ to a single value,
the spectral function is obtained for $\mu=\mu^{\prime}=10^i\;(i=2,\,3,\,4)$ to estimate the systematic error.

\nocite{*}

\bibliography{SpM}

\end{document}